\documentclass[11pt]{JHEP3}
\usepackage{amsmath,amssymb}
\usepackage{epsfig}
\usepackage{dsfont}
\def\x'{\mathaccent 19 x}
\def\y'{\mathaccent 19 y}
\def\n'{\mathaccent 19 n}
\def\u'{\mathaccent 19 u}
\def\et'{\mathaccent 19 \eta}
\def\th'{\mathaccent 19 \theta}
\def\lam'{\mathaccent 19 \lambda}
\def\varet'{\mathaccent 19 \vartheta}
\def\rh'{\mathaccent 19 \rho}
\def\ph'{\mathaccent 19 \Phi}
\def\xb'{\mathaccent 19 {\bar{x}}}

\def\m{\mu}
\def\n{\nu}

\def\TO{\theta_0}

\def\bsp{\be\begin{split}}

\def\d{\delta}
\def\e{\epsilon}
\def\m{\mu}
\def\TI{\theta_1}

\def\ni{\noindent}
\def\wt{\widetilde}

\def\be{\begin{equation}}
\def\ee{\end{equation}}
\newcommand{\bea}{\begin{eqnarray}}
\newcommand{\eea}{\end{eqnarray}}

\def\TI{\theta_1}

\def\a {\alpha}
\def\b {\beta}
\def\s {\sigma}

\def\g {\gamma}

\def\D{\Delta}

\def\e{\epsilon}

\def\Tr{\text{Tr}}
\def\G{\Gamma}

\def\t{\tau}

\def\ni{\noindent}
\def\t{\tau}
\preprint{HU-EP-09/22}
\title{Correlators of supersymmetric Wilson-loops, protected operators and matrix models in ${\cal N}=4$ SYM}
\author{Antonio Bassetto$^{(a)}$, Luca Griguolo$^{(b)}$, Fabrizio Pucci$^{(c)}$, Domenico~Seminara$^{(c)}$,
Shiyamala Thambyahpillai$^{(a)}$ and Donovan Young$^{(d)}$\\
$^{(a)}$ Dipartimento di  Fisica, Universit\`a  di Padova and
INFN Sezione di Padova,\\ Via Marzolo 8, 31131 Padova, Italy \\
$^{(b)}$  Dipartimento di  Fisica, Universit\`a  di Parma and
INFN Gruppo Collegato di Parma, Viale G.P. Usberti 7/A, 43100 Parma, Italy\\
$^{(c)}$ Dipartimento di Fisica, Universit\`a di
Firenze and INFN Sezione di Firenze, Via  G. Sansone 1, 50019 Sesto Fiorentino, Italy\\
$^{(d)}$ Humboldt-Universit\"at zu Berlin, Institut f\"ur Physik,
Newtonstrasse 15, D-12489 Berlin, Germany\\
\email{bassetto@pd.infn.it, griguolo@fis.unipr.it, pucci@fi.infn.it, seminara@fi.infn.it,
shiyamala.thambyahpillai@pd.infn.it, dyoung@physik.hu-berlin.de}}
\abstract{
We study the correlators of a recently discovered family of BPS Wilson loops
in ${\cal N}=4$ supersymmetric $U(N)$ Yang-Mills theory. When the contours
lie on a two-sphere in the space-time, we propose a closed expression
that is valid for all values of the coupling
constant $g$ and for any rank $N$, by exploiting the suspected relation with two-dimensional gauge theories.
We check this formula perturbatively at order ${\cal O}(g^4)$ for two latitude Wilson loops and
we show that, in the limit where one of the loops shrinks to a point, logarithmic
corrections in the shrinking radius are absent at ${\cal O}(g^6)$. This last result strongly supports the
validity of our general expression and suggests the existence
of a peculiar protected local operator arising in the OPE of the Wilson loop. At strong coupling we compare
our result to the string dual of the ${\cal N}=4$ SYM correlator in the limit of large separation,
presenting some preliminary evidence for the agreement.
}
\keywords{}
\begin{document}

\renewcommand{\thefootnote}{\arabic{footnote}}
\setcounter{footnote}{0}

\section{Introduction}
The supersymmetric Maldacena-Wilson \cite{Rey:1998ik,Maldacena:1998im} loops in
${\cal N}=4$ supersymmetric Yang-Mills theory (SYM) were recently
generalized to include a class of contours contained in an $S^3$, which
also include a path-dependent coupling to the scalar fields of the
theory \cite{Drukker:2007qr,Drukker:2007dw}. A subset of those Wilson
loops are contained in a great $S^2$ and their discoverers pointed out
an exact solvability and a potential connection to QCD$_2$
\cite{Drukker:2007qr,Drukker:2007yx}. These loops are given by (we consider our $S^2$ in
hyperplane $x^0=0$)
\be\label{theloop}
W = \frac{1}{N} \Tr {\cal P} \exp \oint d\t \,
\left( i \,\dot x^i A_i + \e_{ijk}\, x^j \dot x^k \,M^i_I \,\Phi_I \right)
\ee
\ni where $x^i(\t)$ (where $i=1,\ldots,3$, $I=1,\ldots,6$) is a closed
path on $S^2$, and $M^i_I$ is a $3\times 6$ matrix satisfying $M
M^T=1$ and which we will take to be $M^i_i=1/R$ (no summation implied and $R$
is the $S^2$ radius)
and all other entries zero. At the level of the vacuum expectation
value (VEV) there is considerable evidence that\footnote{$L_n^m$ is
  the Laguerre polynomial $L_n^m(x)=1/n!\exp[x]x^{-m}(d/dx)^n
  (\exp[-x]x^{n+m})$.}
\be\label{mmVEV}
\langle W \rangle = \frac{1}{N} L_{N-1}^1
\left(-g^2 \,\frac{{\cal A}_1{\cal A}_2}{{\cal A}^2}\right)
\exp \left( g^2 \,\frac{{\cal A}_1{\cal A}_2}{2{\cal A}^2}\right),
\ee
where ${\cal A}_1$ is the area on the sphere enclosed by the Wilson
loop, while ${\cal A} = {\cal A}_1 + {\cal A}_2$ is the total sphere
area. To begin with, the 1/2 BPS circle (given by an equator) has been
proved to be given by (\ref{mmVEV})
\cite{Erickson:2000af,Drukker:2000rr,Pestun:2007rz} and there are
strong arguments in favour of the 1/4 BPS circle of
\cite{Drukker:2006ga} (given by a latitude) also being captured by
(\ref{mmVEV}). At ${\cal O}(g^2)$, (\ref{mmVEV}) was proven for
general contours in \cite{Drukker:2007qr,Drukker:2007dw}.  This result
was further confirmed at ${\cal O}(g^4)$ in
\cite{Bassetto:2008yf,Young:2008ed}.  The significance of the result
is that it agrees with the calculation of the VEV of the Wilson loop in
QCD$_2$ on an $S^2$ in the zero instanton sector \cite{Bassetto:1998sr} with the
couplings related by\footnote{We use different conventions for the Yang-Mills actions in two and
four dimensions that differ by a factor two,  in keeping with the original references on the subject.}
\be
g_{2d}^2 = -\frac{g^2}{\cal A}.
\ee
The idea that a class of ${\cal N}=4$ SYM Wilson loops might be
exactly solvable and equivalent to Wilson loops in a lower dimensional theory
is very attractive, and hints at a relationship between two very different
quantum field theories. More specifically one could infer that the
localization procedure presented in \cite{Pestun:2007rz} could also
apply to this more general class, pointing towards the existence of
a sector of non-local topological observables in ${\cal N}=4$ SYM.
Standard field theoretical arguments should then suggest the presence of
protected local operators arising in the OPE of the Wilson loop (see \cite{Drukker:2009sf}
for related research in this direction).

To substantiate these ideas we need to go beyond the level of the
one-point function of Wilson loops and consider correlators of loops.
A first step in this direction was undertaken in \cite{Young:2008ed},
where a perturbative computation of the correlator of two latitudes at
order ${\cal O}(g^6)$ was undertaken. Lacking a zero-instanton QCD$_2$
result to compare to, in \cite{Young:2008ed} the generalization to
$S^2$ of the Wu-Mandelstam-Leibbrandt (WML)
\cite{Wu:1977hi,Mandelstam:1982cb,Leibbrandt:1983pj} prescription for
QCD$_2$ in the plane proposed in \cite{Drukker:2007qr,Drukker:2007dw}
was used. Indeed, this prescription has been recently shown to be
equivalent to the zero-instanton QCD$_2$ result
\cite{Giombi:2009ds}\footnote{A disagreement was erroneously present in
\cite{Young:2008ed}.}.

In the present paper we derive a general formula for correlators of
BPS Wilson loops with arbitrary contours on $S^2$ in terms of the
multi-matrix model governing the zero instanton expansion of QCD$_2$.
The result is valid for any coupling constant $g$ and for any value of
$N$: we compute explicitly the matrix integral for the correlator of
two loops. Our general expression survives a series of non-trivial
tests. First of all we calculate in ${\cal N}=4$ perturbation theory
the correlator of two latitude Wilson loops at ${\cal O}(g^4)$,
finding perfect agreement with the matrix model result. Next we
provide compact formulas for the perturbative ${\cal O}(g^6)$
contribution, generalizing the results of \cite{Bassetto:2008yf}, from
which a numerical evaluation can be easily performed (we will report
on this point in the future \cite{toappear}). Here we prefer instead
to investigate analytically the limit where one of the two latitudes
shrinks to zero size: because our nonperturbative formula is an order
by order polynomial in the shrinking radius, the absence of
logarithmic terms is a crucial test of the matrix representation.  We
find indeed the absence of leading logarithms in the shrinking radius,
a quite non-trivial result, differing dramatically from the analogous
computation of non-BPS correlators \cite{Arutyunov:2001hs} where logs
are present.

Interestingly, by analyzing the OPE of the shrinking Wilson loop one can relate the
absence of the logarithmic terms to the protection of a local operator
which may be expressed as the trace of the square of a twisted field
strength. Work
\cite{Drukker:2009sf} concerning super-protected local operators
could be extended to
also include this novel operator, which is based on very similar
symmetries. We discuss this issue in section \ref{sec:twist}.

Armed with our general result we can therefore take the large $N$ and strong coupling
limit and try to compare it to the ${\cal N}=4$ correlator from the string side. In the
limit that the two latitudes shrink to opposite poles on the sphere,
this calculation reduces to the semi-classical exchange of
supergravity (SUGRA) modes between the two string worldsheets
describing the Wilson loops at strong coupling. We find that at
leading order in the large-separation limit, the matrix model result seems
to capture the exchange of the SUGRA modes dual to a certain
chiral primary operator. Other modes, dual to other protected
operators present in the weak coupling OPE, should also be carefully included
to test definitively this result at strong coupling.
We find moreover an intriguing pattern of matching
between the QCD$_2$ result and the exchange of heavier modes dual to chiral primary operators of
higher dimension, which seems to extend to arbitrary order in the
large-separation expansion. We have not yet understood the meaning of
this highly non-trivial pattern of matching.

In this paper we present a survey of our investigations, deferring a
complete
analysis with all the relevant technical details to a future publication.

{\it Note added}: as this manuscript was being completed \cite{Giombi:2009ms} appeared,
presenting a partial overlap with the results of this paper.

\section{Symmetries of the loops and of their correlators}
\label{sec:twist}
We start by considering ${\cal N}=4$ SYM Wilson loops that are a special case
of the general construction presented in \cite{Drukker:2007qr,Drukker:2007dw}.
They are 1/4 BPS supersymmetric loops with the contour defined on a latitude of $S^2$,
first put forward in \cite{Drukker:2006ga}. Writing the Wilson loop as
\be\label{latwl}
W = \frac{1}{N} \Tr {\cal P} \exp \oint d\t \,
\left( i \,\dot x^\mu A_\mu + |\dot x| \,\Theta^I \,\Phi_I \right),
\ee
\ni the latitudes are given by the following closed paths on an $S^2
\subset \mathbb{R}^4$ and on another $S^2 \subset S^5$ which gives the coupling
to the scalar fields $\Phi_I$ ($\mu = 1,\ldots,4$, $I=1,\ldots,6$),
\begin{equation}
x^\mu = R\,(\sin\TO\cos\t,\sin\TO\sin\t,\cos\TO,\,0),\ \ \ \
\Theta^I = (-\cos\TO\cos\t,-\cos\TO\sin\t,\sin\TO,\,0,\,0,\,0)\nonumber.
\end{equation}
\FIGURE[ht]{
\includegraphics*[bb= 46 46 205 205,height=1.5in]{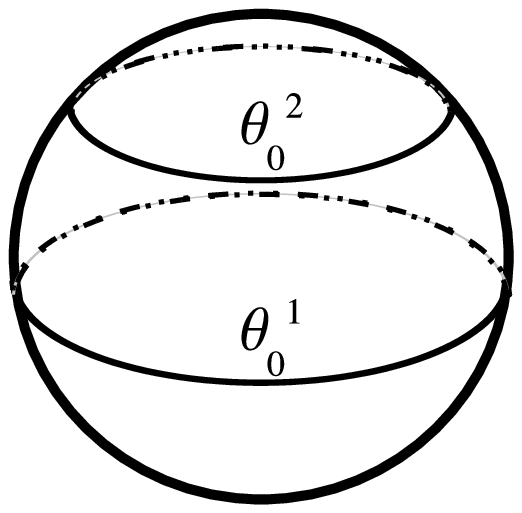}
\caption{\label{fig:twolats} Two Wilson loops given by latitudes at polar angles
  $\TO^1$ and $\TO^2$.}
}
Two such Wilson loops are pictured in figure \ref{fig:twolats}. The
supersymmetries preserved by these operators are fully described in
\cite{Drukker:2007qr}, see section 2.3.1: here we just repeat some details
of that analysis which are relevant to our work.

Under general superconformal transformations we have for the ${\cal N}=4$ SYM bosons
\be
\d_\e A_\m = \bar \Psi \g_\m \,\e,\qquad
\d_\e \Phi_i = \bar\Psi \G_i \,\e,\quad
\e = \e_0 + x^\m \g_\m \e_1.
\ee
Demanding that $\d_\e W = 0$ one finds two relations
\bsp\label{latsusy}
&\g_{12} \,\e_1 = - \G_{12}\, \e_1,\\
&\G_3\, \e_0 = \Bigl[ i\g_{12} +\cos\TO\, \g_3 \G_2 (\g_{23} + \G_{23}) \Bigr]\e_1.
\end{split}
\ee
It is clear that each of them reduce the supersymmetry by half, and therefore a single
latitude is 1/4 BPS. We will be mainly interested in the correlator of two such
Wilson loops, as shown in figure \ref{fig:twolats}. The first relation
in (\ref{latsusy}) is shared between two such latitudes, whereas the
second is clearly not. Thus two latitudes are collectively 1/8 BPS,
each sharing half of their individual supersymmetry. The same reasoning applies
of course to a collection of $n$ latitudes, resulting always in a 1/8 BPS system.

\subsection{Operator product expansion}

In the next section we will present results of a perturbative
calculation of the correlator of two latitudes and, in particular, we will consider
the limit where one of the latitudes shrinks to a point at the pole of the sphere. The emerging structure
can be usefully understood in terms of the OPE and its physical meaning is quite transparent.

The crucial observation is that, viewed from a comparably large distance, the unshrunken Wilson loop
sees the shrunken loop as a collection of local operators
\cite{Berenstein:1998ij}: the quantum behavior is encoded into Wilson coefficients and anomalous dimensions.
The story was worked out in detail for two circular Wilson-Maldacena loops in
\cite{Arutyunov:2001hs}. Here, for the 1/4 BPS latitude, we will find
that the relevant OPE is quite different, giving rise to novel operators which appear to
have protected dimensions.

When analysing the OPE, we can in fact consider the general situation of loops with $arbitrary$ contours on $S^2$ that are
generically 1/8 BPS. As noticed in \cite{Drukker:2007qr} the Wilson loop (\ref{theloop}) can be written in terms of a
new gauge connection
\be
 {\cal A}_i=A_i+i\e_{ijk}\, x^j   \,\frac{\Phi^k}{R}.
\ee
The OPE expansion will appear particularly simple using this generalized connection\footnote{We thank Nadav Drukker for suggesting this course of investigation to us.}. The first step is to determine the classical expansion of our Wilson loops in terms of local gauge-invariant operators when the circuit is small.
To achieve this goal we shall assume that the circuit can be written as follows
\be
x^i(t)=x_0^i+r \hat x^i(t),
\ee
$x_0$ being the point about which the loop is shrinking and $r$ a parameter that will control the limit.
We expand the contour integral
by exploiting the Fock-Schwinger gauge $(x-x_0)^i {\cal A}_{i}(x)=0$, where the following formula holds in terms of the new gauge curvature
${\cal F}_{ji}$
\be
{\cal A}_i(x)=\int_0^1 d\lambda \lambda (x-x_0)^j {\cal F}_{ji}( x_0+\lambda (x-x_0)).
\ee
The leading order result is given by
\be
\begin{split}
\oint_C dt {\cal A}_i(x)\dot{x}^i=&\frac{r^2}{2} {\cal F}_{ij}( x_0)\oint_C dt\hat{x}^i(t) \dot{\hat{x}}(t) +O(r^3)
= \frac{r^2}{2}\e^{ijk}{\cal F}_{ij}( x_0) n_k(x_0) +O(r^3),
\end{split}
\ee
$n_i(x_0)$ being a normal vector to $S^2$ at the point $x_0$, depending on $x_0$ and the contour. The expansion could of course be extended to any
given order in $r$, producing a series of local operators $O^{J}_C(x)$ determined by the particular shape of the Wilson loop, the
generalized connection  ${\cal A}_i$ itself depending on the contour. Because these operators should share the
BPS properties of the associated Wilson loop, we obtain a practical realization of the
proposal of \cite{Drukker:2009sf}: in particular we could expect that their correlation functions,
when restricted to the relevant $S^2$, are somehow protected from quantum corrections. This would imply severe
constraints on Wilson loop correlators. Let us exemplify the consequences for latitude correlators (we will consider here for
simplicity the $SU(N)$ case).

In our specific example we take as our shrinking point the north pole,  $x_0=R (0,0,1)$, while $r=\sin\theta_0$ and
$\hat x^i(t)=R(\cos t,\sin t,\tan\frac{\theta_0}{2})$. Due to the trace in the path-exponential the
first non-vanishing contribution to the OPE is quadratic in the fields, and we get explicitly at leading order
\be\label{OPEN}
W_0 = 1 + \frac{\pi^2 \sin\theta_0^4}{2 N} \, {\cal O}_{\cal F}(x_0)
\ee
where
\be\label{theop}
\mathcal{O}_{\cal F}(x_0) =\Tr \Bigl[ 2 R\,\Phi_3 - i R^2\,F_{12} - R^2\,(\partial_1 \Phi_1 + \partial_2
  \Phi_2) \Bigr]^2.
\ee
We note a peculiar feature that makes this OPE quite different from the usual
circular Wilson-Maldacena case \cite{Arutyunov:2001hs}: operators of classical dimension 2, 3, and 4
all couple with the same power of the parameter which sets the size of the shrinking latitude:
the polar angle $\theta$ (in the standard case the power is the classical dimension itself).
Indeed the overall scale $R$ of the $S^2$ is just a place keeper. The
conformality of ${\cal N}=4$ SYM prevents it from playing any r\^ole, and
it drops out of the calculation of any observable.

We notice that we can easily obtain the leading
term of the two latitude correlator at order $g^4$ from the OPE (\ref{OPEN}), once we restore the canonical
normalization for the fields. We just need to compute the correlation function
\be
\langle {\cal O}_{\cal F}(x_0)\oint dt \,
\left( \,\dot x_1^i A_i(x_1) -i \e_{ijk}\, x_1^j \dot x_1^k \,\Phi^i(x_1) \right)\rangle=i\frac{\cos \left(\theta _1\right)+1}{4 \pi },
\ee
that enters in the Wick contraction. Taking the relevant color traces we get
\be
\label{OPEA}
\frac{\langle W_0 W_1\rangle}{\langle W_0\rangle \langle W_1\rangle}-1=\frac{g^4}{8} \left(2\sin^2\frac{\theta_0}{2} \cos^2\frac{\theta_1}{2}\right)^2=\frac{g^4 r^4}{32 }\cos^2\frac{\theta_1}{2}.
\ee
The above result will be confirmed in the next section by the finite size correlator. Actually we can learn something more:
the general expectation for the structure of the OPE of a shrinking Wilson loop is given by
\cite{Arutyunov:2001hs,Berenstein:1998ij,Semenoff:2001xp}
\be
\frac{W}{\langle W \rangle} = 1 + \sum_{J} \, \xi_J(g^2) \, L^{\D_J}\,
O_J(x)
\ee
where $L$ is the size of the shrinking loop, and $O_J(x)$ is an operator
of classical dimension $J$ and quantum dimension $\D_J = J +
g^2 \,\D_J^{(1)} + \ldots$. The Wilson coefficients $\xi_J(g^2)$ depend on the
coupling constant $g^2$. The curious structure of the latitude OPE is
a reflection of the fact that the coefficients $\xi_J(g^2)$ which
describe the coupling of the Wilson loop to a specific operator
$O_J(x)$ are themselves functions of $\theta$ \cite{Semenoff:2006am},
and can be expanded as $\xi_J(g^2,\theta) = \sum_k \xi^{(k)}_J(g^2) \,
\theta^{k}$ in the limit $\theta\to 0$. This provides us with the general
structure for the OPE of $W_0$
\bsp\label{OPE}
\frac{W_0}{\langle W_0 \rangle} &= 1 + \sum_{J} \, \xi_J(g^2,\theta_0) \, \theta_0^{\D_J}\,
O_J(x_0)= 1 + \sum_{J,\,k} \, \xi^{(k)}_J(g^2) \, \theta_0^{\D_J+k}\,
O_J(x_0)=\\
& = 1 + \xi^{(2)}_2 \, \theta_0^{\D_2 + 2}\, O_2(x_0)
 + \xi^{(1)}_3 \, \theta_0^{\D_3 + 1}\, O_3(x_0) + \xi^{(0)}_4\,\theta_0^{\D_4}\,O_4(x_0) +
 \ldots\,,
\end{split}
\ee
where we have dropped the scale $R$ (to restore it replace
$O_J(x_0) \to R^{\D_J}\,O_J(x_0)$), and have noted the vanishing of
$\xi_2^{(0,1)}$ and $\xi_3^{(0)}$ from the explicit expression of (\ref{theop}). The explicit form
of $O_{2,3,4}(x_0)$ is simply obtained from ${\cal O}_{\cal F}(x_0)$. Actually there
are multiple operators of the same classical dimension, so there
is an extra suppressed index on the $\xi_J(g^2,\theta_0)$, $\D_J$, and
$O_J(x_0)$, which is implicitly summed over in (\ref{OPE}). In the last
line we are referring only to the operators appearing in (\ref{theop}) as these are the only ones present at leading order in $\theta_0$. We derive the following general relation
in the shrinking limit
\bsp
\frac{\langle W_1 \, W_0 \rangle}{\langle W_1 \rangle \langle W_0 \rangle}
=  &1 + \xi^{(2)}_2 \, \theta_0^{\D_2 + 2}\, \langle W_1 \,O_2(x_0)\rangle
 + \xi^{(1)}_3 \, \theta_0^{\D_3 + 1}\,\langle W_1 \, O_3(x_0) \rangle + \xi^{(0)}_4\,\theta_0^{\D_4}\,\langle W_1 \,O_4(x_0)\rangle +
 \ldots\,.
\end{split}
\ee
We notice that when expanded at small coupling the $\theta_0^{\D_J}$ terms generically produce
 logarithms $\theta_0^{\D_J} = \theta_0^J+ g^2 \,\D_J^{(1)}\,\theta_0^J \log
\theta_0 +\ldots$ if quantum corrections modify the classical dimensions. The quantities $\xi^{(2)}_2$, $\xi^{(1)}_3$, and
$\xi^{(0)}_4$ may easily be read-off in our case from (\ref{theop}). Since the operators appearing in the
explicit expression are quadratic in the fields, one has that $\xi^{(2)}_2$, $\xi^{(1)}_3$, and
$\xi^{(0)}_4$ lead as $g^4$. We therefore generally expect terms of
the form $g^6 \log \theta_0$ to show up in the perturbative expansion of the correlator at order
$g^6$, in the shrinking limit.

The presence of logarithmic corrections would be a signal that anomalous
dimensions are playing a part, suggesting that the full interacting theory should be taken into account and localization techniques
would not be sufficient in the exact computation. It would also rule out the relation with two-dimensional Yang-Mills that
produces just polynomial dependence on $\theta$ at any order of perturbation theory, as we will see in section \ref{sec:qcd2}.
In section \ref{sec:pert} we show that, surprisingly, no such logarithmic
terms appear at order $g^6$, supporting the matrix model proposal. This indicates that the composite operator ${\cal O}(x)$, arising
from the OPE of the BPS loops \eqref{theloop}, should be protected - at least at the first non-trivial quantum order. In
other words logarithmic divergences should be absent in the two-point function $\langle \, {\cal O}(x_1)\,{\cal O}(x_2)\,\rangle$,
when $x_{1,2}$ belong to the relevant $S^2$, in the same way as the operators defined in \cite{Drukker:2009sf}. It is not
difficult to show in fact that ${\cal O}(x)$ inherits the BPS properties of the latitude loop, and a certain amount of
supersymmetry is preserved by its correlators.

\section{Perturbative results on Wilson loop correlators}
\FIGURE[ht]{
\epsfig{file=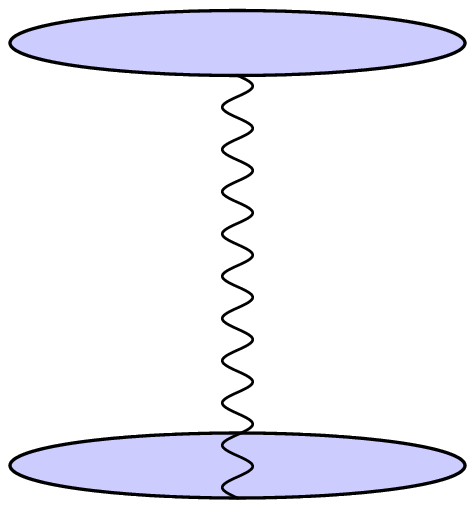,width=20mm,height=20mm}
\caption{\label{single} $g^2-$diagram}
}

\label{sec:pert}
In this section we perform a perturbative analysis up to order\footnote{Only at this order do
the interactions start contributing to the connected Greens functions.} $g^6$ for the connected correlator
$\mathcal{W}(C_1,C_2)\equiv W(C_1,C_2)-W(C_1) W(C_2)$ of two
latitudes in the case that the gauge group is $U(N)$. To begin with, we shall consider the  $g^2$ diagram
 depicted in fig. \ref{single}. [Notice that this contribution would be absent in a $SU(N)$theory.]

\noindent
In order to carry out the computation, we parameterize the two circuits
using polar coordinates
\begin{equation}
\begin{split}
 C_1=&R (\sin\theta_1\cos \tau,\sin\theta_1\sin \tau,\cos\theta_1)\\
  C_2=&R (\sin\theta_2\cos \sigma,\sin\theta_2\sin \sigma,\cos\theta_2),
\end{split}
\end{equation}
and  define the effective propagator $\Delta_{C_1 C_2}(\tau,\sigma)$ connecting the two loops
\be
\begin{split}
\Delta_{C_1 C_2}(\tau,\sigma)=&\frac{2}{N}\langle\Tr(\mathcal{A})(\tau) \Tr(\mathcal{A})(\sigma)\rangle_0=
-\frac{\sin \theta _1 \sin \theta _2\left(\cos \left(\tau-\sigma\right) \left(\cos \theta _1 \cos \theta_2-1\right)+\sin \theta _1
 \sin \theta _2\right)}{8 \pi ^2 \left(\cos \theta _1 \cos \theta _2+\cos
   \left(\tau-\sigma\right) \sin \theta _1 \sin \theta _2-1\right)},
\end{split}
\ee
where  $\mathcal{A}$ denotes the effective field $i A_\mu(x)\dot{x}^\mu+\! \Theta_I \Phi^I(x)|\dot{x}|$. Then the $g^2-$contribution is
given by
\be
\left.\mathcal{W}(C_1,C_2)\right|_{g^2}
=\frac{g^2}{2 N} \int_{0}^{2\pi} d\tau d\sigma~~ \Delta_{C_1 C_2}(\tau,\sigma)=\frac{\lambda}{N^2} \frac{A_1 A_2}{ A^2} \ \ \ \ \ \ \  (\lambda\equiv g^2N),
\ee
where $A$ is the total area of the sphere, and $A_1$ and $A_2$ are the areas enclosed by the two Wilson-loops given by
\be
\frac{A_1}{A}=\frac{2\pi (1-\cos\theta_1)}{4\pi}=\sin^2\frac{\theta_1}{2}\ \ \ \
\frac{A_2}{A}=\frac{2\pi (1+\cos\theta_2)}{4\pi}=\cos^2\frac{\theta_2}{2}.
\ee
At order $g^4$, we have  to consider the diagrams in fig. \ref{gto4}. First, we shall consider the
contribution $S_{g^2-g^2}$ due to diagram $(b_1)$. Its evaluation reduces to the following integral over
the circuits
\FIGURE[ht]{
\epsfig{file=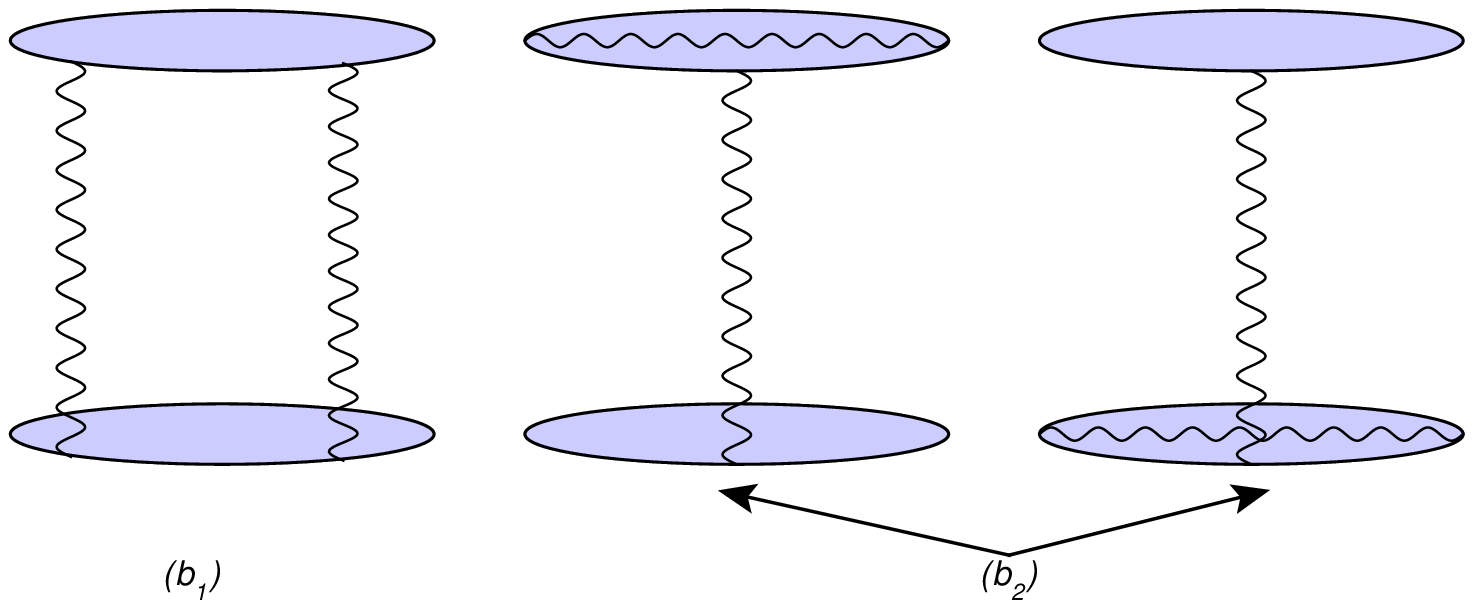,width=98mm,height=26mm}
\caption{ \label{gto4} $g^4$ diagrams}
}
\begin{align}
\label{k1}
 S_{g^2-g^2}=&\frac{g^4 }{16} \int_{0}^{2\pi}\!\!\!\!\!d\tau_1  d\tau_2  d\sigma_1 d\sigma_2\mbox{\small$
\biggl[\Delta_{C_1 C_2}(\tau_1,\sigma_1)\Delta_{C_1 C_2}(\tau_2,\sigma_2)+\Delta_{C_1 C_2}(\tau_1,\sigma_2)\Delta_{C_1 C_2}(\tau_2,\sigma_1)\biggr]$}=\nonumber\\
=&\frac{g^4 }{8}\left[ \int_{0}^{2\pi}\!\!\!d\tau_1  d\sigma_1  \Delta_{C_1 C_2}(\tau_1,\sigma_1)\right]^2
=\frac{g^4 }{8}\left[ 2\sin^2\frac{\theta_1}{2}\cos^2\frac{\theta_2}{2}\right]^2=\frac{\lambda^2}{2 N^2} \frac{A_1^2 A_2^2}{ A^4}.
\end{align}
Next we shall consider the contribution $S_{g-g^3}$ due to the  two diagrams $(b_2)$.   The sum of the two
diagrams yields
\begin{align}
S_{g-g^3}=&\frac{g^4 }{4!} \oint_{C_1}\!\! d\tau_1 \oint_{C_2}\!\! d\sigma_1  d\sigma_2  d\sigma_3
\mbox{\small$
\left(\Delta_{C_1 C_2}(\tau_1,\sigma_1)\Delta_{C_2 C_2}(\sigma_2,\sigma_3)+\Delta_{C_1 C_2}(\tau_1,\sigma_2)\Delta_{C_2 C_2}(\sigma_1,\sigma_3)+\right.$}\nonumber\\
&\mbox{\small$
\left.+\Delta_{C_1 C_2}(\tau_1,\sigma_3)\Delta_{C_2 C_2}(\sigma_1,\sigma_2)\right)$}+ (C_1\leftrightarrow C_2)=\nonumber
\\
=&\frac{g^4}{16} \left(2\sin^2\frac{\theta_1}{2}\cos^2\frac{\theta_2}{2}\right) (\sin ^2\theta_1+\sin ^2\theta_2)
=\frac{\lambda^2}{2 N^2 A^4} A_1 A_2( A_1 A_3+A_2 A_3+2 A_1 A_2),
\end{align}
where
$
\Delta_{C_2 C_2}(\sigma_i,\sigma_j)=\frac{\sin ^2\theta_2}{8 \pi ^2}
$  and $A_3=A-A_1-A_2$.
If we sum all the contributions at order $g^4$, the total result is
\FIGURE[l]{
\includegraphics[width=50mm,height=22mm]{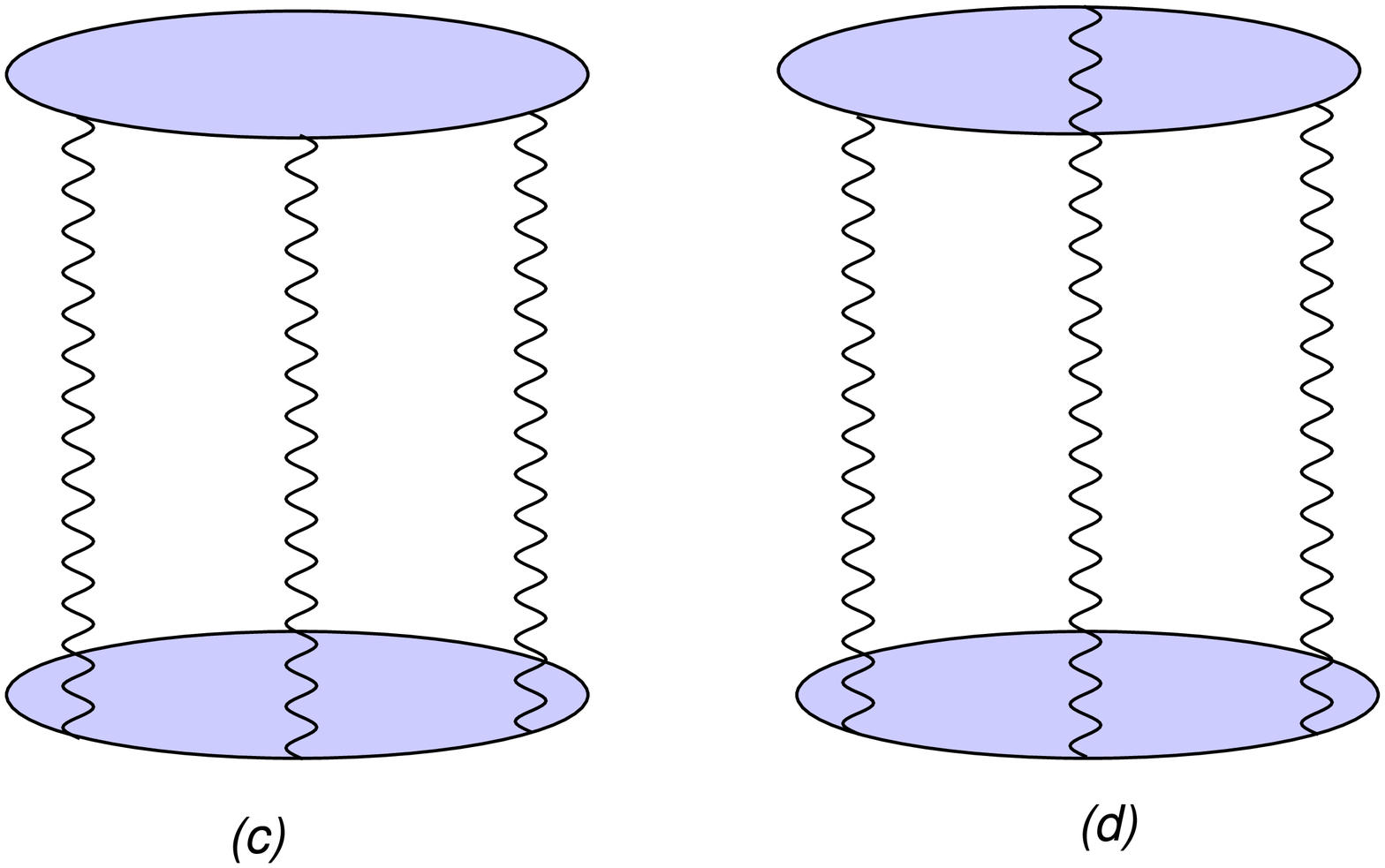}
\caption{\label{ladder} Triple-exchange: (a) planar diagram; (b) non-planar diagram.}
}
\be
\left.\mathcal{W}(C_1,C_2)\right|_{g^4}
=\frac{\lambda^2 }{2N^2 A^4 }A_1 A_2({A}_1 {A}_3+{A}_2 {A}_3+3 {A}_1 {A}_2).
\ee
A remark on the $S_{g^2-g^2}$ contribution is in order. This is the only contribution
in a $SU(N)$ theory and one can verify that its small $r-$expansion is in agreement
with  the OPE result [\ref{OPEA}], supporting the idea that the leading contribution
to the Wilson-loop is determined only by $\mathcal{O}_{\cal F}$.

\noindent
\FIGURE[l]{
\epsfig{file=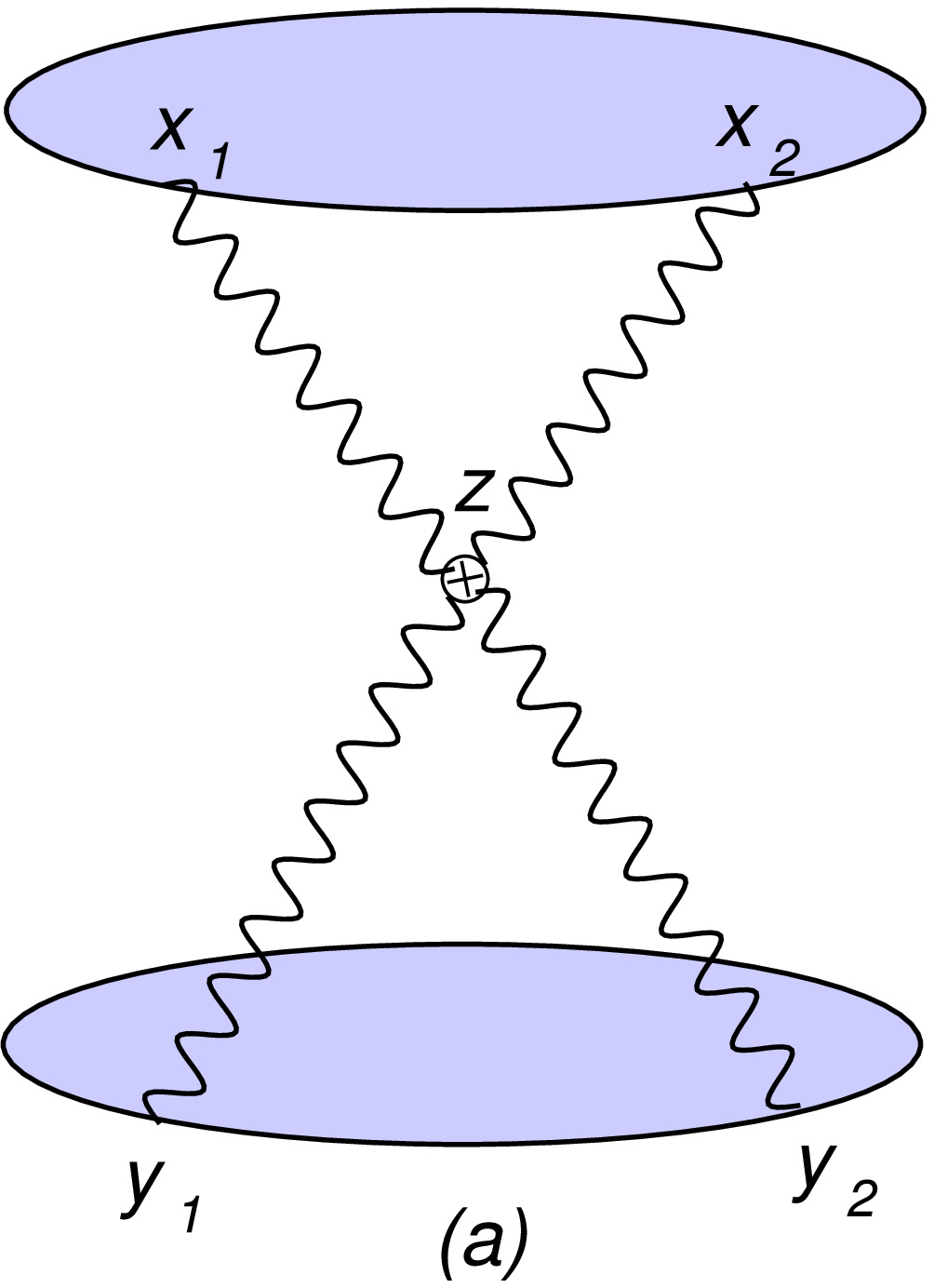,width=25mm,height=25mm}
\epsfig{file=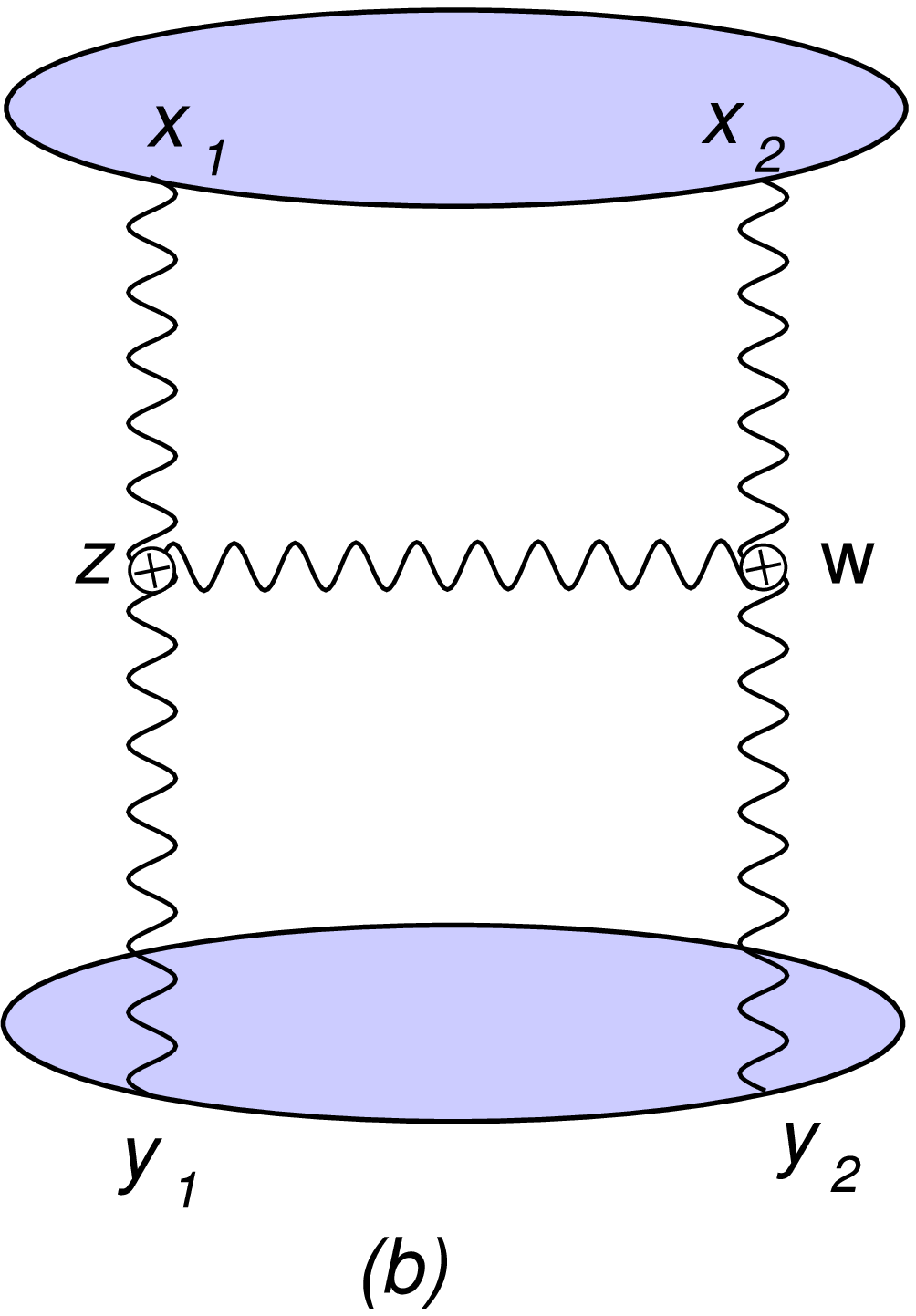,width=25mm,height=25mm}
\caption{\label{X-H} (a) X-diagram; (b) H-diagram.}
}
We now come to considering the $g^6$ contribution. Since, at this order,
the $\mathcal{N}=4$ interactions will start contributing, a complete analytic
evaluation of all the relevant integrals  is out of reach. However one can
write compact formulas which can be used as a starting point for a numerical
evaluation \cite{toappear}. We shall exploit this  possibility in a future paper. Here
we shall instead be interested in singling out the coefficients of
contributions of the form $r^k\log(r)$,  potentially present in the evaluation
of the connected  correlator. The knowledge of these coefficients already provides non trivial information on the properties of the correlator. In fact, as explained
in the previous section, a  non-vanishing  result for these coefficients  would clash
with the expectation that the correlator  localizes.\hfill

\noindent
For this computation, we limit our attention to the  gauge group $SU(N)$ and we can separate
the  diagrams into two classes: the ladder diagrams
and the interaction diagrams.  The ladder diagrams are depicted in fig. \ref{ladder} and it is easy to realize
that they cannot generate any contribution of the form $r^k \log(r)$. They are actually analytic in the small
$r-$limit.  The contributions $r^k \log(r)$ are instead generated by the interactions diagrams in figs.
\ref{X-H} and  \ref{IY}. The origin of this non analytic behavior
can be traced back to the small distance singularities appearing in the integration over the position
of the vertices. Thus in order to  extract  these logarithmic singularities,  we have to first perform
these integrations analytically, and only after that can  we expand in powers
of the radius. To illustrate the procedure let us start by considering the $X$-diagram. Its expression
can be cast into the following compact form
\be
\label{X}
\begin{split}
\mathbf{X}=\frac{\lambda^3}{8 N^2}&\int_0^{2\pi}\!\!\!\!\! d\tau_1d\tau_2d\sigma_1 d\sigma_2
\left[(\dot{x}_1\circ\dot{y}_2)(\dot{x}_2\circ\dot{y}_1)-\right.\\
&\left.-(\dot{x}_1\circ\dot{x}_2)(\dot{y}_1\circ\dot{y}_2)\right]
\mathcal{I}^{(4)}(x_1,x_2,y_1,y_2),
\end{split}
\ee
where $(\dot{x}\circ\dot{y})=\dot{x}\cdot\dot{y}-|\dot{x}| |\dot{y}| \Theta_{\dot{x}}\cdot \Theta_{\dot{y}}$
with $|\dot x| \Theta_{\dot x}^I =M_I^i\epsilon_{irs} \dot x^r x^s$ and
\be
\label{A4}
\mathcal{I}^{(4)}(x_1,x_2,y_1,y_2)\equiv\frac{1}{(2\pi)^8} \int \frac{d^4 w}{(x_1-w)^2(x_2-w)^2(y_1-w)^2(y_2-w)^2}.
\ee
Here and in the following $x_i\equiv x(\tau_i)$ and $y_i\equiv y(\sigma_i)$ will denote points on the upper and
lower latitudes respectively (see fig. \ref{fig:twolats}).
The integration over $w$ in \eqref{A4} can be performed and it is then straightforward to extract the singular  part
when we shrink the latitude $\theta=\theta_1$ to the north-pole of the sphere $S^2$ (see appendix $A$ for details.)
The singular part is given by
\begin{align}
&\mathcal{I}^{(4) sing.}(x_1,x_2,y_1,y_2)=-\frac{{ \log r }}{128\pi^6}\times\nonumber\\
&\times
\!\int_0^1\!\!
  \frac{d\alpha}{\mbox{\small $(1-\alpha){( y_1- x_2)^2(y_2 -x_1)^2}-\alpha(1-\alpha) {( x_1- x_2)^2(y_1-y_2)^2}+
 {\alpha( y_1- x_1)^2(y_2-x_2)^2}$}},
\end{align}
where $r=\sin\theta_1$.   The integration over the circuit is straightforward and can be evaluated by Taylor-expanding
in $r$. At leading order we find that
\be
\mathbf{X^{sing}}=\frac{5 r^4 \cos ^4\left(\frac{\theta _2}{2}\right) \log (r)}{768 \pi ^2}+O(r^5).
\ee
\noindent
Consider now the $H-$diagram in fig. \eqref{X-H}. We can write the contribution from this diagram as follows
{\small
\be
\label{eq11w}
\begin{split}
\mathbf{H}
=&-\frac{\lambda^3}{8 N^2}\int d^4 w\left [\underset{\mathbf{A}_1}{{P}^M(x_1, y_1, w)\square_w {P}^M (x_2,y_2,w)}+
\underset{\mathbf{B}_1}{{P}^M(x_1, y_1, w)\square_w {Q}^M (x_2,y_2,w)}+\right.\\
&\ \ \ \ \ \ \ \ \ \ \ +\left.\underset{\mathbf{B}_2}{{Q}^M(x_1, y_1, w)\square_w {P}^M (x_2,y_2,w)}+\underset{\mathbf{A_2}}{
{Q}^M(x_1, y_1, w)\square_w {Q}^M (x_2,y_2,w)}\right],
\end{split}
\ee
}
where
\be
\label{cubo1}
P^M(x_i,y_i,w)=\int_0^{2\pi}\!\!\!\!\!\! d\tau_i  d\sigma_i \left[2 \dot {y_i}^M (\dot x_i\cdot \partial_{{y_i}}
\mathcal{I}_i(x_i,y_i,w))-2 \dot x_i^M (\dot {y_i}\cdot\partial_{x_i}\mathcal{I}_i(x_i,y_i,w)
)\right]
\ee
and
\be
\label{cubo2}
Q^M(x_i,y_i,w)=
\int_0^{2\pi}\!\!\!\!\!\! d\tau_i d\sigma_i
(\dot{x}_i\circ\dot{y}_i)(\partial_{x^M_i}\mathcal{I}_i(x_i,y_i,w)-\partial_{y^M_i}\mathcal{I}_i(x_i,y_i,w)).
\ee
In eqs. \eqref{eq11w}, \eqref{cubo1} and \eqref{cubo2},
the index $M$ is a ten-dimensional label running from $1$ to $10$ and in particular we have defined $x^M\equiv (x^\mu, i \Theta^I |\dot x|)$
and $\partial_M\equiv (\partial_\mu,0)$. The function $\mathcal{I}_{1}(x_i,y_i,w)$ is defined by the scalar integral
\be
\label{II1}
\mathcal{I}_{1}(x_i,y_i,w)=\frac{1}{(2\pi)^6}\int  \frac{d^4 z}{(x_i-z)^2(y_i-z)^2(w-z)^2}.
\ee
The spatial components $P^\mu$ of $P^M$ satisfy the following  two simple identities: $z_\mu P^\mu=\partial_\mu P^\mu=0$, as
can easily be checked by direct computation. Moreover, for two latitudes
parallel to the plane $(2,3)$, $P^1$ and
$P^4$ trivially vanish. Since $P^\mu$ is a just a function of $z^\mu$,  all these properties are consistent if and only if $P^\mu=0$.
This result simplifies dramatically the computation for the correlator of two latitudes: in fact the contributions  $\mathbf{B_1}$
and $\mathbf{B_2}$ in  \eqref{eq11w}  are identically zero. Recall, in fact, that $Q^M$ is different from zero (by construction) only when
$M$ is spatial. Thus we are just left with $\mathbf{A_1}$ and $\mathbf{A_2}$ to be computed.

\noindent
Let us first compute first $\mathbf{A_2}$. It is convenient to rewrite this contribution as follows
\be
\mathbf{A_2}=\frac{\lambda^3}{8 N^2}\int_0^{2\pi}\!\!\!\!\!\! d\tau_{1}d\tau_{2} d\sigma_1 d\sigma_2
\dot x_1\circ \dot {y_1} \dot x_2\circ \dot {y_2}
( \partial_{x_1}-\partial_{y_1})\cdot (\partial_{x_2}-\partial_{y_2})\mathcal{H}(x_1,y_1;x_2, y_2),
\ee
where
\be
\begin{split}
\mathcal{H}(x_1, y_1; x_2, y_2)=&\frac{1}{(2\pi)^{10}}
\int \frac{ d^{4} z d^{4}w}{(x_1-z)^2(y_1- z)^2 (z-w)^2 (x_2-w)^2 (y_2-w)^2}.
\end{split}
\ee
The action of $(\partial_{x_1}-\partial_{y_1})\cdot  (\partial_{x_2}-\partial_{y_2})$  on  $\mathcal{H}(x_1, y_1; x_2, y_2)$ can then be evaluated with the identity (A.7) given in \cite{Beisert:2002bb}. One finds
\be
\label{yy}
\begin{split}
 (\partial_{x_1}-\partial_{y_1})\cdot& (\partial_{x_2}-\partial_{y_2}) \mathcal{H}(x_1, y_1; x_2, y_2)=\\
=& \frac{1}{(x_1-y_1)^2 (x_2-y_2)^2}\left[\mathcal{I}^{(4)} (x_1,y_1,x_2,y_2)((x_1-x_2)^2 (y_1-y_2)^2-(x_1-y_2)^2(x_2-y_1)^2)+\right.\\
&+\left.\frac{1}{(2\pi)^2}(Y(x_1,x_2,y_2)-Y(y_1,x_2,y_2)+Y(x_2,x_1,y_1)-Y(y_2,x_1,y_1))\right],
\end{split}
\ee
where
$
{Y} (x_1,x_2,x_3)\equiv{\mathcal{I}_1 (x_1,x_2,x_3)}[(x_1-x_3)^2-(x_1-x_2)^2].
$
When the first latitude ($\theta=\theta_1$) is shrunk to zero the logarithmically divergent terms can be generated
by $\mathcal{I}^{(4)}$ and by the $Y$ that  depends both on $x_1$ and $x_2$.  Therefore we can write
\be
\label{A2}
\begin{split}
\mathbf{A_2^{sing.}}=&\frac{\lambda^3}{8 N^2}\int_0^{2\pi}\!\!\!\!\!\! d\tau_{1}d\tau_{2} d\sigma_1 d\sigma_2
\frac{(\dot x_1\circ \dot {y_1})(\dot x_2\circ \dot {y_2})}{(x_1-y_1)^2 (x_2-y_2)^2 }\biggl[\mathcal{I}^{(4)sing.} (x_1,y_1,x_2,y_2)((x_1-x_2)^2 (y_1-y_2)^2-\\
&-(x_1-y_2)^2(x_2-y_1)^2)+\frac{1}{(2\pi)^2}(Y^{sing.}(x_1,x_2,y_2)+Y^{sing.}(x_2,x_1,y_1))\biggr],
\end{split}
\ee
where we have defined
\[
{Y}^{sing.} (x_1,x_2,y_2)\equiv{\mathcal{I}_1^{sing.} (x_1,x_2,y_2)}[(x_1-y_2)^2-(x_1-x_2)^2]
\]
and
\[
{Y}^{sing.} (x_2,x_1,y_2)\equiv{\mathcal{I}_1^{sing.} (x_1,x_2,y_2)}[(x_2-y_2)^2-(x_2-x_1)^2].
\]
The expression for  $\mathcal{I}_1^{sing.}$ is given in appendix A. The  integration  over the circuits
can then be easily performed with the help of Mathematica if we first expand the integrand of  \eqref{A2}
in powers of $r$. At leading order we find
\be
\mathbf{A_2^{sing.}}=-\frac{7 r^4 \cos ^4\left(\frac{\theta _2}{2}\right) \log (r)}{1536 \pi ^2}+O(r^5).
\ee
To complete the evaluation of the $H-$diagram we have to compute the contribution $\mathbf{A_1}$.  The first step is to add
two total derivatives to the integrand of $P^M$
{\small
\be
\begin{split}
P^M(x_1,y_1,w)=&\int_0^{2\pi}\!\!\!\!\!\! d\tau_{1} \int^{2\pi}_0\!\!\!\!\!\! d\sigma_1 \left[2 \dot {y_1}^M (\underbrace{\dot x_1\cdot \partial_{{y_1}}
\mathcal{I}_1(y_1-w,x_1-w)-\dot x_1\cdot \partial_{{x_1}}\mathcal{I}_2(y_1-w,x_1-w)}_{K_1})-\right.\\
&-\left. 2\dot x_1^M (\underbrace{ \dot {y_1}\cdot\partial_{x_1}\mathcal{I}_1(x_1-w,y_1-w)-\dot {y_1}\cdot\partial_{y_1}\mathcal{I}_2(x_1-w,y_1-w))
}_{K_2})\right].
\end{split}
\ee
}
These two new terms obviously yield a vanishing result when the integration runs along the circuits. The function $\mathcal{I}_2(x,y)$ is defined in
appendix $A$. Since  the following identity for $\mathcal{I}_1$ and $\mathcal{I}_2$ holds \cite{Bassetto:2008yf}
\be
\begin{split}
\frac{\partial}{\partial x^\mu}\mathcal{I}_1(x,y)-\frac{\partial}{\partial y^\mu}\mathcal{I}_2(x,y)=
-\frac{ 1 }{32\pi^{4}} \frac{x^\mu}{x^2}\frac{\log\left(\frac{(x-y)^2}{y^2}\right)}{[(x-y)^2-y^2]},
\end{split}
\ee
the combination $K_1$  appearing in $P^M$ can be rearranged in the following compact form
\be
\label{U1}
\begin{split}
K_1=&-\frac{1}{64\pi^4(y_1-w)^2}\frac{d}{d\tau_1} \left[\mathrm{Li}_2\left(1-\frac{(x_1-y_1)^2}{(x_1-w)^2}\right)+\frac{1}{2}
\left(\log\left[\frac{(x_1-w)^2}{(x_2-y_2)^2}\right]\right)^2\right]+\\
&+\frac{1}{32\pi^4}\frac{(x_1-w)\cdot\dot{x}_1}{(x_1-w)^2(y_1-w)^2}\log\left(\frac{(x_1-y_1)^2}{(x_2-y_2)^2}\right).
\end{split}
\ee
The combination  $K_2$ can be also recast into the same form. The only difference from \eqref{U1} is that the roles of  $x_1$ and $y_1$, and of
$\tau_1$ and $\sigma_1$, are
exchanged.
The terms in $K_1$ and $K_2$ that are total derivatives with respect to $\tau_1$ and $\sigma_1$ can be dropped
since they yield a vanishing contribution to  $P^M$, and we are left with the  compact expression
\be
\label{cubo}
\begin{split}
P^M(x_1,y_1,w)=\frac{1}{16\pi^4} &\int_0^{2\pi}\!\!\!\!\!\! d\tau_{1}  d\sigma_1
\frac{\dot {y_1}^M (x_1-w)\cdot\dot{x}_1-\dot x_1^M (y_1-w)\cdot\dot{y}_1}{(x_1-w)^2(y_1-w)^2}\log\left(\frac{(x_1-y_1)^2}{(x_2-y_2)^2}\right).
\end{split}
\ee
Then, if we  take into account that
\be
\begin{split}
-\square_w P^M(x_2,y_2,w)
=&\int_0^{2\pi}\!\!\!\!\!\! d\tau_{1}  d\sigma_1 \left[2 \dot {y_2}^M \dot x_2\cdot \partial_{{y_2}}
- 2 \dot x_2^M \dot {y_2}\cdot\partial_{x_2}\right]\frac{1}{(2\pi)^4} \frac{1}{(x_2-w)^2 (y_2-w)^2},
\end{split}
\ee
we can rewrite the $\mathbf{A_1}$  contribution in the following form
\begin{align}
\label{A1}
\mathbf{A_1}&=\frac{\lambda^3}{4 N^2}\int_0^{2\pi}\!\!\!\!\!\! d\tau_1 d\tau_2 d\sigma_1 d\sigma_2
\log\left(\frac{(x_1-y_1)^2}{(x_2-y_2)^2}\right)\Biggl[ [(\dot {y_1}\circ \dot y_2)  \dot x_2\cdot\partial_{y_2} -
(\dot {y_1}\circ \dot x_2)  \dot y_2\cdot\partial_{x_2}]\times\nonumber\\
&\times \dot x_1\cdot S(x_1, x_2,y_1,y_2)- [(\dot {x_1}\circ \dot y_2)  \dot x_2\cdot\partial_{y_2} -
(\dot {x_1}\circ \dot x_2)  \dot y_2\cdot\partial_{x_2}]\dot y_1\cdot S( x_1,x_2,y_1,y_2)\Biggr]
\end{align}
where
\be
\label{AA}
S^\mu(x_1, x_2,y_1,y_2)\equiv-\frac{1}{(4\pi^2)^4}\int d^4 w \frac{w^\mu}{(x_1-w)^2(y_1-w)^2(x_2-w)^2 (y_2-w)^2}.
\ee
The nice feature of \eqref{A1} is the disappearance  of one of the integrations over the position of the vertices. Although
this result simplifies the procedure for extracting the logarithmic terms appearing in the limit $\theta_1\to 0$,
the  computation is still a little bit cumbersome and some of the details are given in appendix $A$. Here we shall only give the
final result after the integration over the circuits. At the leading order in $r(\equiv \sin\theta_1)$, we find
\be
\mathbf{A_1}=\frac{r^4 \cos ^4\left(\frac{\theta _2}{2}\right) \log (r)}{512 \pi ^2}
\ee
The final set of diagrams to compute are depicted in fig. \ref{IY}.  We have two contributions that we call respectively $\mathbf{IY}_{\mathbf{up}}$
[(c) in fig. \ref{IY}] and $\mathbf{IY}_{\mathbf{down}}$  [(d) in fig. \ref{IY}], and a diagram which takes into account the one-loop correction to the effective propagator [(e) in  fig. \ref{IY}]. We
shall denote this third diagram by $\mathbf{Budiag}$.  To begin with we focus our attention on $\mathbf{IY}_{\mathbf{up}}$, whose
expression is
\be
\begin{split}
\mathbf{IY}_{\mathbf{up}}=\frac{\lambda^3J}{8 N^2}\int^{2\pi}_0 \!\!\!\!\!\!d\tau_1 d\tau_2 d\tau_3 d\sigma_2 \varepsilon(\tau_1,\tau_2,\tau_3)
\{ (\dot{x}_1\circ \dot{y}_2 )\dot{x}_2\cdot(\partial_{y_2}-\partial_{x_1})-(\dot{x}_1\circ \dot x_2) \dot{y}_2 \cdot\partial_{x_2}\} \mathcal{I}_{1}(x_1,x_2,y_2),
\end{split}
\ee
\FIGURE[ht]{
\includegraphics[height=35mm,width=65mm]{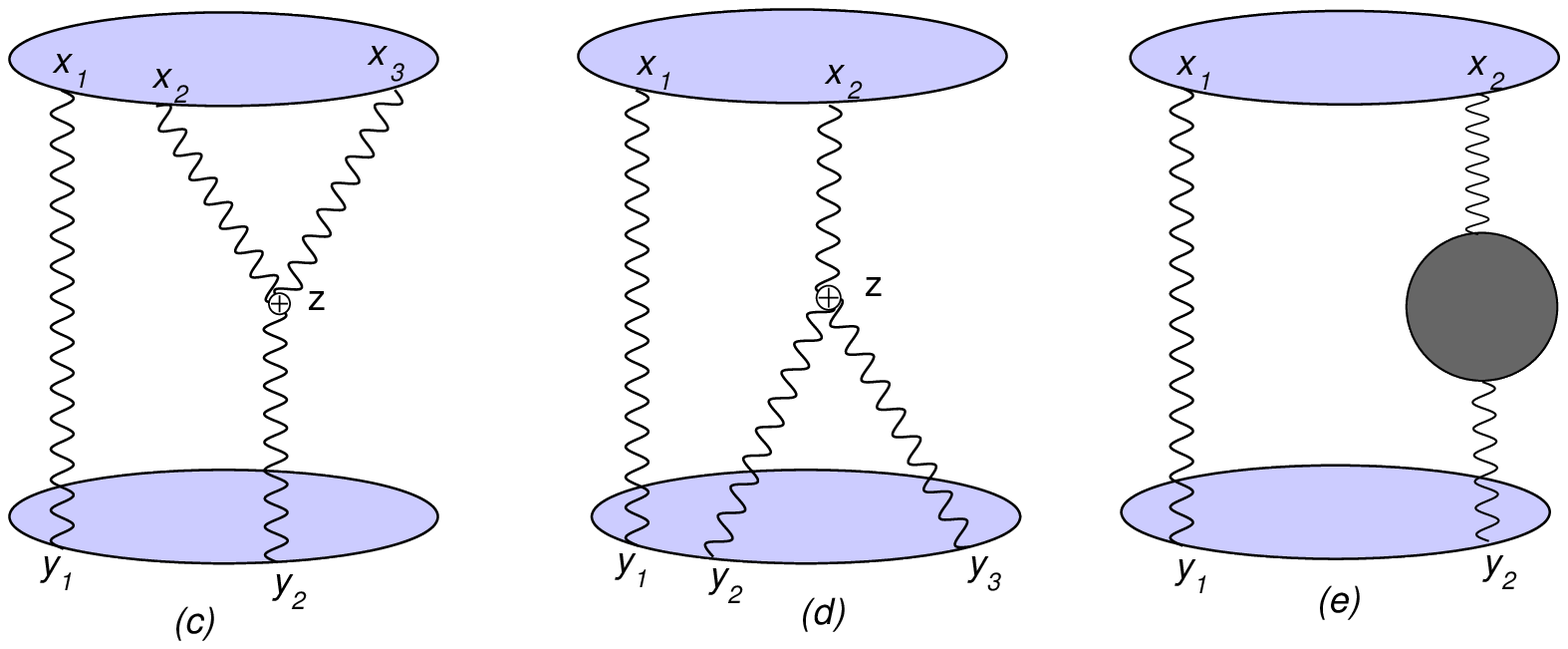}
\caption{\label{IY} The two {\it ``IY-diagrams''} and the self-energy correction.}
}\noindent
and on $\mathbf{IY}_{\mathbf{down}}$, which is obtained from $\mathbf{IY}_{\mathbf{up}}$ by exchanging the roles of  $\sigma$ and $\tau$
(and therefore $x_i$ and $y_i$).
Here $J$ is the constant defined by the integral\footnote{This integral is independent of $\tau_i$, namely it is constant, because
the integrand is function only of  $\sigma_1-\tau_i$ and we are integrating a periodic function over the interval $[0,2\pi]$.}
\be
J= \int_0^{2\pi}d\sigma_1 (\dot{x}_i\circ \dot y_1)D(x_i-y_1),
\ee
where $D(x)$ is the usual Feyman propagator.
When we shrink the upper circle to a point, the logarithmic behavior can originate only from $\mathbf{IY}_{\mathbf{up}}$. The contribution $\mathbf{IY}_{\mathbf{down}}$ yields analogous behavior when we shrink the lower circle.
However, when evaluating $\mathbf{IY}_{\mathbf{up}}$, we also encounter  \textit{divergences at coincident points} ($\tau_1\to\tau_2$)
in the integration
over the upper circuit.  This singularity though is compensated by the  standard ultraviolet-divergence of the  self-energy
graph: half of diagram  $\mathbf{Budiag}$ cancels the divergence for $\tau_1\to \tau_2$, while the other half cancels the same singularity
in $\mathbf{IY}_{\mathbf{down}}$ for $\sigma_1 \to \sigma_2$. Therefore,  in order to safely extract the logarithmic behavior when
we shrink the circuit to zero, we have to first realize this cancellation.

\noindent To begin with, performing a trivial integration by parts, we can rewrite  $ \mathbf{IY}_{\mathbf{up}}$ in the following form
\be
\begin{split}
\mathbf{IY}_{\mathbf{up}}
=&\frac{\lambda^3 J}{8 N^3}\left[\int^{2\pi}_0 d\tau_1 d\tau_2 d\tau_3 d\sigma_2 \varepsilon(\tau_1,\tau_2,\tau_3)
\right.\{ (\dot{x}_1\circ \dot{y}_2 )2\dot{x}_2\cdot\partial_{y_2}-(\dot{x}_1\circ \dot x_2) \dot{y}_2 \cdot\partial_{x_2}\} \mathcal{I}_{1}(x_1,x_2,y_2)+\\
&-2\int^{2\pi}_0 d\tau_1 d\tau_2 d\sigma_2
(\dot{x}_1\circ \dot{y}_2 ) \mathcal{I}_{1}(x_1,x_2,y_2)\left.+\frac{1}{2}\int^{2\pi}_0 d\tau_1  d\tau_3 d\sigma_2
(\dot{x}_1\circ \dot{y}_2 ) \mathcal{I}_{1}(x_1,x_1, y_2)
\right].
\end{split}
\ee
The singular part for coincident points is now singled out in the last term, which is proportional to $\mathcal{I}_{1}(x_1,x_1, y_2)$.
Since
\be
\begin{split}
\mathbf{Budiag}=-&\frac{\lambda^3 J }{8  N^2}\int_0^{2\pi} d\tau_1  d\tau_3
 d\sigma_2 (\dot{x}_1\circ\dot{y}_2)\mathcal{I}_{1}(x_1,x_1, y_2).
\end{split}
\ee
 half of $\mathbf{Budiag}$ exactly cancels the singularity present in $\mathbf{Y}_{{\mathbf{up}}}$ and we  are left with
\be
\begin{split}
\mathbf{IY}_{\mathbf{up}}
=&\frac{\lambda^3 J}{8N^2 }\left[\int^{2\pi}_0 d\tau_1 d\tau_2 d\tau_3 d\sigma_2 \varepsilon(\tau_1,\tau_2,\tau_3)
\right.
\{ (\dot{x}_1\circ \dot{y}_2 )2\dot{x}_2\cdot\partial_{y_2}-(\dot{x}_1\circ \dot x_2) \dot{y}_2 \cdot\partial_{x_2}\} \mathcal{I}_{1}(x_1,x_2,y_2)-\\
&\left.-2\int^{2\pi}_0 d\tau_1 d\tau_2 d\sigma_2
(\dot{x}_1\circ \dot{y}_2 )\mathcal{I}_{1}(x_1,x_2,y_2)\right].
\end{split}
\ee
This expression does not exhibit any singularity at coincident points.
The logarithmic part arising when we shrink the upper circle to a point is then obtained by replacing $\mathcal{I}_1$ in the above expression
with the $\mathcal{I}_1^{sing.}$ found in appendix A. Next we Taylor-expand in $r$  and integrate over the circuits. At
leading order in $r$ we find
\be
\mathbf{IY}_{\mathbf{up}}^{\mathbf{sing.}}=-\frac{r^4 \cos ^4\left(\frac{\theta _2}{2}\right) \log (r)}{256 \pi ^2}+O(r^5)
\ee
Let us now sum all the different contributions at leading order in $r$
\be
\mathbf{X}^{\mathbf{sing.}}+\mathbf{IY}_{up}^{\mathbf{sing.}}+\mathbf{A_1}^{\mathbf{sing.}}+\mathbf{A_2}^{\mathbf{sing.}}=\frac{r^4 \cos ^4\left(\frac{\theta _2}{2}\right) \log (r)}{ \pi ^2}\left(\frac{5}{768}-\frac{1}{256}+\frac{1}{512}-\frac{7}{1536}\right)=0\ !
\ee
Namely, we have verified that the logarithmic singularities cancel at the first non trivial order. This implies that the effective anomalous dimension of
the operator $\mathcal{O}_{\cal F}$ defined in the previous section vanishes at one-loop, supporting the idea that this operator is actually protected.

As we will show in the next section, this result is consistent with
the result coming from the zero instanton expansion of QCD$_2$.

\section{The conjectured matrix model description}
\label{sec:qcd2}

\noindent
In the previous sections we have tried to argue  that the  correlator of two (or more) Wilson-loops of type
\eqref{theloop}
might be an exactly  solvable quantity since it belongs to a topological sector of $\mathcal{N}=4$. This idea, in fact,
passes a certain number of non trivial tests: [a]
the observable is $1/8$ BPS independently of the position and the form of the loops \cite{Drukker:2007yx};
[b] there is a {\it candidate} topological twist of the $\mathcal{N}=4$ theory, where one of the supercharges preserving the correlator
becomes a scalar \cite{Drukker:2007yx}; [c] finally, if we  compute the behavior of the correlator when one of the circuits shrinks to a point we
get a smooth limit with no logarithmic singularity. This last property in particular, should be contrasted with what happens for
the correlator of two circular Maldacena-Wilson loops \cite{Arutyunov:2001hs}: there the logarithmically singular behavior was present
and signaled the impossibility of a matrix model description for this observable \cite{Arutyunov:2001hs}.\hfill\\
\FIGURE[ht]{
\epsfig{file=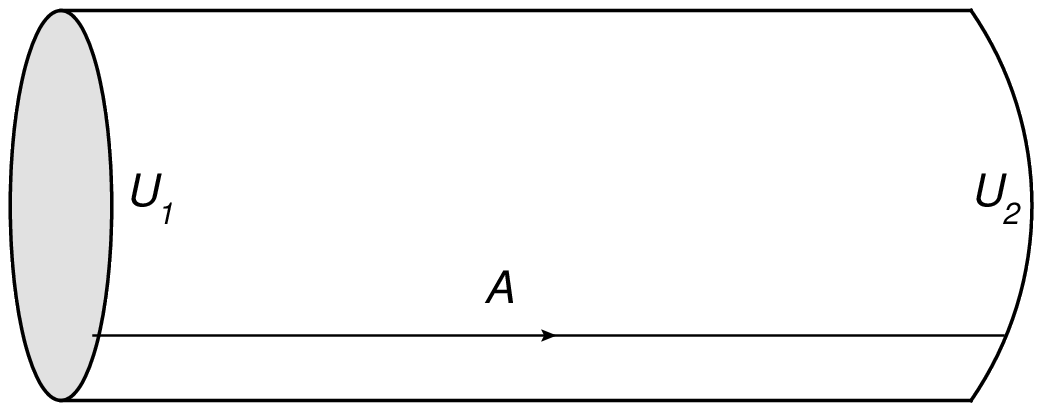,width=40mm}
\caption{\label{propagator} Cylinder amplitude}
}
\noindent
In this section we shall accept this idea, and focus our attention on the problem of writing  a general formula for the
correlator of two Wilson-loops. The starting point is to recall that the expectation value of one Wilson-loop appears to be
computed by the  matrix model describing the  zero-instanton sector of a Wilson loop for $QCD_2$ on the two sphere
\cite{Drukker:2007yx,Bassetto:2008yf,Young:2008ed}. Since the single Wilson loop and the correlator generically share
the same symmetries we expect that this equivalence  also extends to the case of correlators. Therefore we conjecture that
the correlator of two Wilson loops of type \eqref{theloop} is given by the multi-matrix model, which evaluates
the zero-instanton sector of the correlator of two loops for $QCD_2$ on $S^2$.

\noindent
The construction of this matrix model is quite simple since $QCD_2$ is an almost topological theory (it is invariant under
area-preserving diffeomorphisms) and its observables can be computed with the help of some simple {\it string-like} Feynman-rules
\cite{Witten:1992xu}. For the present computation we need just three ingredients: the cylinder amplitude (\textit{heat-kernel propagator}),
the disc and the Feynman rule for the observable, {\it i.e. the Wilson loop}. The first quantity  is represented in fig. \ref{propagator} and is given by \hfill
\be
\label{cc}
\mathcal{K}(A; U_1, U_2)=\langle U_2 |  e^{-\frac{g^2 A\triangle}{2}}|
U_1\rangle =  \sum_{R} \chi_R( U_1) \chi^\dagger_R(U_2) e^{-\frac{g^2 A}{2} C_2(R)},
\ee
where $A$ is the area of the cylinder and the sum runs over all the representations $R$ of $U(N)$. The amplitude also depends on the
two holonomies $U_1$ and $U_2$ defined on the  two borders of the cylinder.  There is in fact a dual representation for the cylinder amplitude where
the sum over representations is replaced with a sum over the instanton charges
\be
\mathcal{K}(A; U_1, U_2)=\!\sum_{P\in S_N} \frac{(g^2 A)^{-N/2}}{J(\theta_i)J(\phi_i)} \sum_{\mathbf{\ell}\in \mathds{Z}^N } (-1)^{P+(N-1)\sum \ell_i}
\exp\left({-\frac{1}{2 g^2 A} \sum_{i=1}^N (\phi_i-\theta_{P(i)}+2\pi i \ell_i)^2}\right),
\ee
where  $\{e^{i\theta_i}\}$  and $\{e^{i\phi_i}\}$ are the eigenvalues of the matrices $U_1$ and $U_2$ respectively and
\[
J(\theta_i)=\prod_{i\le j} 2 \sin\left(\frac{\theta_i-\theta_j}{2}\right).
\]
The disc is obtained from \eqref{cc} by choosing one of the two
holonomies to be  trivial - namely equal to the identity. Finally, the insertion of a Wilson loop with winding number $n$
is realized by introducing the factor
$ \mathrm{Tr}(U^n)$ at the border of the cylinder. The  amplitude for the correlator of two non-intersecting loops with winding numbers
$n_1$ and $n_2$ is schematically represented
in fig. \ref{amplitude}, and the corresponding expression is given by the following two-matrix integral over the unitary matrices:
\FIGURE[hb]{
\epsfig{file=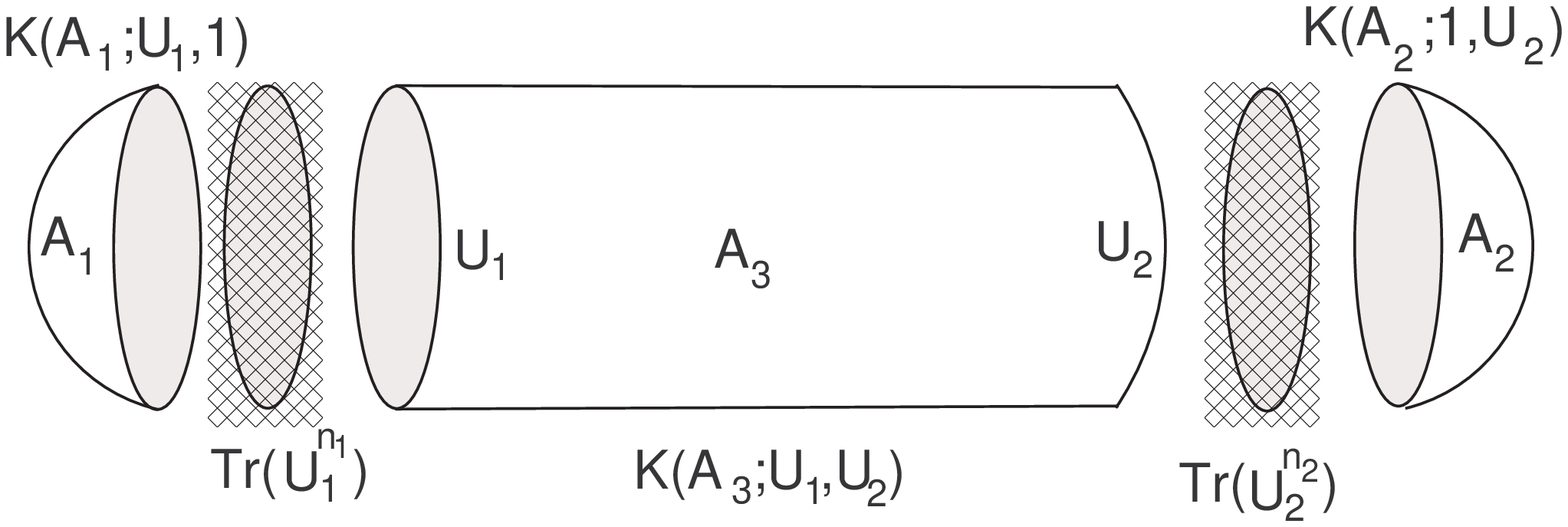,width=100mm}
\caption{\label{amplitude} The string-like Feynman-diagram for the correlator of two Wilson-loops.}
}
\be
\label{amp1}
\!\!\!\!\!
\begin{split}
\tilde{\mathcal{W}}(A_1,A_2)=&\frac{1}{N^2}\!\int\!\! \mathcal{D} U_1\mathcal{D} U_2 \Tr(U_1^{n_1})
\Tr(U_2^{n_2})
\mathcal{K}(A_1; \mathds{1},U_1) \mathcal{K}(A_3; U_1, U_2) \mathcal{K}(A_2; U_2,\mathds{1})=\\
=&\frac{1}{N^2}\sum_{P\in S_N}\sum_{\ell,m,s\in \mathds{Z^N}}\int d^N\theta d^N\phi J^2(\theta_i) J^2(\phi_i)\left(\sum_{r,s=1}^N
e^{i n_1\theta_r+i n_2 \phi_s}\right)\times\\
&\times \frac{(g^2 A_1)^{-\frac{N^2}2}}{J(\theta_i)}  (-1)^{(N-1)\sum_{i}\ell_i}\Delta(\theta_i+2\pi
\ell_i)
\exp\left(-\frac{1}{2 g^2 A_1} \sum_{i=1}^N (\theta_i+2\pi \ell_i)^2\right)\times\\
&\times
\! \frac{(g^2 A_3)^{-N/2}}{J(\theta_i)J(\phi_i)}  (-1)^{P+(N-1)\sum s_i}
\exp\left({-\frac{1}{2 g^2 A_3} \sum_{i=1}^N (\phi_i-\theta_{P(i)}+2\pi i s_i)^2}\right)\\
&\times \frac{(g^2 A_2)^{-\frac{N^2}2}}{J(\phi_i)}  (-1)^{(N-1)\sum_{j}m_j}\Delta(\phi_j+2\pi
m_j)
\exp\left(-\frac{1}{2 g^2 A_2} \sum_{i=1}^N (\phi_i+2\pi m_i)^2\right),
\end{split}
\ee
$\Delta$ being the Vandermonde determinant.
The amplitude $\tilde{\mathcal{W}}(A_1,A_2)$ is related to the true correlator by the relation
$\tilde{\mathcal{W}}(A_1,A_2)=\mathcal{Z}\mathcal{W}(A_1,A_2)$, where $\mathcal{Z}$ is the partition function
of $QCD_2$ on the sphere.
We can extend the region of integration over the entire $\mathds{R}^{2N}$ by means of the sum over $\ell$ and $m$
and we can rewrite the above expression as
\be
\label{PP}
\begin{split}
&\tilde{\mathcal{W}}(A_1,A_2)
=\frac{(g^4 A_1 A_2)^{-\frac{N^2}2}(g^2 A_3)^{-\frac{N}2}}{N^2}\sum_{P\in S_N}\sum_{s\in \mathds{Z^N}}
(-1)^{P+(N-1)\sum s_i}\!\!\!\int_{\mathds{R}^{2N}}   \!\!\!\!\!\!\! d^N\theta d^N\phi  \left(\sum_{r,s=1}^N
e^{i n_1\theta_r+i n_2 \phi_s}\right)\!\!\times\\
&\times\Delta(\theta_i)\Delta(\phi_i)
\exp\left(-\frac{1}{2 g^2 A_1} \sum_{i=1}^N \theta_i^2-\frac{1}{2 g^2 A_3} \sum_{i=1}^N (\phi_i-\theta_{P(i)}+2\pi i s_i)^2
-\frac{1}{2 g^2 A_2} \sum_{i=1}^N \phi_i^2\right).
\end{split}
\ee
The result \eqref{PP} is the exact amplitude and it contains all instantonic corrections. To single out the zero-instanton
sector of this amplitude it is sufficient  to consider  the case where all instanton numbers $s_i$ vanish.
If we introduce the diagonal matrices $\Theta=\mathrm{diag}(\theta_1,\dots,\theta_N)$ and $\Phi=\mathrm{diag}(\phi_1,\dots,\phi_N)$, using the Itzykson-Zuber integration formula and defining the hermitian matrices  $V_1=U^{-1}\Theta U $ and $V_2=V\Phi V^{-1}$, we can recast the original
integral as the following hermitian two matrix model for the correlator of two Wilson loops\footnote{
The generalization of this result to the case of $n$ loops is trivial
\[
\begin{split}
\mathcal{W}(A_1,\dots,A_n)
=&\frac{1}{C_N N^2}\int D V_1\dots D V_n {\rm e}^{-\sum_{i=1,n}\frac{1}{2g^2 A_i}\mathrm{Tr}(V_i^2)-\sum_{j=1}^{n-1}\frac{1}{2 g^2 A_{j, j+1}}\mathrm{Tr}((V_j- V_{j+1})^2)}
\mathrm{Tr}(e^{i r_1 V_1})\cdots\mathrm{Tr}(e^{i r_n V_n}),
\end{split}
\]
where $A_1,A_n$ are the areas enclosed respectively by the first and and last loop (by "enclosed" we mean the region of $S^2$ not containing other loops) and $A_{i,i+1}$ is the area between the $i-$th and $(i+1)-$th loop.
}
\begin{align}
W(A_1,A_2)=&\frac{1}{C_N N^2}\int D V_1 D V_2 {\rm e}^{-\frac{A_1+A_3}{2g^2 A_1 A_3}\mathrm{Tr}(V_1^2)-\frac{A_2+A_3}{2g^2 A_2 A_3}\mathrm{Tr}(V_2^2)+\frac{1}{g^2 A_3}\mathrm{Tr}(V_1 V_2)}
\mathrm{Tr}(e^{i n_1 V_1})\mathrm{Tr}(e^{i n_2 V_2})=\nonumber
\\
\label{orth}
=&\frac{1}{C_N N^2}\int D V_1 D V_2 {\rm e}^{-\frac{1}{2g^2 A_1}\mathrm{Tr}(V_1^2)-\frac{1}{2g^2 A_2}\mathrm{Tr}(V_2^2)-\frac{1}{2 g^2 A_3}\mathrm{Tr}((V_1- V_2)^2)}
\mathrm{Tr}(e^{i n_1 V_1})\mathrm{Tr}(e^{i n_2 V_2}),
\end{align}
where the normalization is chosen to be
\be
\label{nor}
C_N= \int D V_1 D V_2 {\rm e}^{-\frac{A_1+A_3}{2g^2 A_1 A_3}\mathrm{Tr}(V_1^2)-\frac{A_2+A_3}{2g^2 A_2 A_3}\mathrm{Tr}(V_2^2)+\frac{1}{g^2 A_3}\mathrm{Tr}(V_1 V_2)}.
\ee
Actually, in the sector $s_i=0$ of \eqref{PP}, the angular integration can be
performed by means of an expansion in terms of Hermite polynomials
 and by exploiting the relation between
integrals over Hermite polynomials and Laguerre polynomials.
Then one finds the following finite $N$ closed expression for the connected correlator
\begin{align}
\label{fin}
&W(A_1,A_2)-W(A_1) W(A_2)=\nonumber\\
=&\frac{1}{N^2 } e^{-\frac{\left(A_1 A_2 (n_1
 +n_2)^2+A_3 (n^2_1 A_1+n_2^2 A_2)\right) g^2}{2
   A}}
  L^{1}_{N-1}\left({\frac{g^2 \left(A_3 n_1 +A_2(n_1+ n_2) \right)\left(A_1 (n_1+n_2) +A_3 n_2\right)}{A}} \right)+\nonumber\\
  &-\frac{1}{N^2} e^{-\frac{\left(A_1 \left(A_2+A_3\right) n_1^2+A_2 \left(A_1+A_3\right) n_2^2\right) g^2}{2
  A}}\times\\
  &\times\sum_{i_1,i_2=1}^N   \left(-\frac{g^2 n_1 n_2 A_1 A_2  }{A}\right)^{i_2-i_1}
  \frac{(i_1-1)! }{(i_2-1)!} L_{i_1-1}^{i_2-i_1}\left({\frac{g^2 n_2^2 A_2(A_3+A_1)}{ A}}\right)
L^{i_2-i_1}_{i_1-1}\left({\frac{g^2 n_1^2 A_1(A_3+A_2)}{ A}}\right)\!\!,\nonumber
 \end{align}
where $A=A_1+A_2+A_3$ is the total area of the sphere. For small $g$ this expression can be expanded in a power series and one finds
\be
\label{bbb}
\begin{split}
W(A_1,A_2)&-W(A_1) W(A_2)=-\frac{{A_1} {A_2} g^2n_1 n_2}{ N A}+\\
&+\frac{{A_1} {A_2} ({A_1} {A_2}(n_1^2+n_2^2+n_1 n_2)+{A_3} ({A_1} n_1^2+{A_2}n_2^2)) g^4 n_1 n_2}{2
   A^2}+\\
   &- g^6n_1 {n_2}\left(\frac{{A_1}^3 {A_2} ({A_2}+{A_3})^2 \left(2 N^3+N\right) n_1^4}{24 A^3 N^2}+\frac{{A_1}^3 {A_2}^2
   ({A_2}+{A_3}) \left(2 N^3+N\right) {n_2} n_1^3}{12 A^3 N^2}+\right.\\
   &+\left.\frac{{A_1}^2 {A_2}^2 \left(3 {A_3} ({A_2}+{A_3}) N^2+{A_1}
   \left(3 {A_3} N^2+{A_2} \left(4 N^2+1\right)\right)\right) {n_2}^2 n_1^2}{12 A^3 N}+\right.\\
   &+\frac{{A_1}^2 {A_2}^3 ({A_1}+{A_3}) \left(2
   N^3+N\right) {n_2}^3 n_1}{12 A^3 N^2}+\left.\frac{{A_1} {A_2}^3 ({A_1}+{A_3})^2 \left(2 N^3+N\right) {n_2}^4}{24 A^3 N^2}\right)+O(g^7).
\end{split}
\ee
This result, after decompactifying the sphere, agrees with the perturbative results we have obtained up to
 ${\cal O}(g^6)$ from Feynman graph calculations using the Mandelstam-Leibbrandt prescription for the
vector propagator in light-cone coordinates \cite{toappear}. Let us compare the perturbative result \eqref{bbb}
with the actual computation in $\mathcal{N}=4$ done in section 3. After performing the standard redefinition
$g^2\mapsto -g^2/A$ and setting $n_1=n_2=1$, we find complete agreement up to order $g^4$. Notice,
 moreover, that the agreement with $QCD_2$ demands the absence of logarithmic singularities when the area of
one of the loops is small, to all orders in perturbation theory. Our $g^6$ result of sec. 3 is consistent with this prediction.

\noindent
In order to analyse the large $N$ limit, we can write a simple compact representation for the connected correlator in $\mathcal{N}=4$ SYM
by exploiting a contour representation of the Laguerre polynomials
\begin{align}
\label{ciop2a}
W(A_1,A_2)-W(A_1) W(A_2)=\frac{n_1 n_2}{N^2}\int_{C_1} \frac{dw_1}{2\pi i} \int_{C_2} \frac{dw_2}{2\pi i}
\frac{e^{w_1+w_2+\frac{\lambda  \left(\tilde{A_1} A_1 w_2 n_1^2+\tilde{A_2} A_2 n_2^2 w_1\right)}{A^2 w_2 w_1}} \tilde{A_2} A_1}{\left(\tilde{A_2} n_2 w_1-A_1 n_1
   w_2\right)^2},
 \end{align}
where $\tilde A_1=A-A_1$ and $\tilde A_2=A-A_2$.
This expression  can be computed as an infinite series of Bessel functions. We limit our attention to the case $n_1=n_2=1$ and
are actually interested in  the normalized correlator, which is given by
\be\label{WWmm}
\frac{W_{conn.}}{W_1 W_2}=\frac{\lambda}{N^2 A^2} \tilde{A}_1\tilde{A}_2\sum_{k=1}^\infty
  k  \left (\sqrt{\frac{A_1 A_2}{\tilde{A}_1 \tilde{A}_2} }\right)^{k+1} \frac{I_{k}\left(2{\sqrt{\frac{\lambda A_2 \tilde{A}_2}{A^2} {}}}\right)}{I_{1}\left(2{\sqrt{\frac{\lambda A_2 \tilde{A}_2}{A^2} {}}}\right)}
   \frac{I_{k}\left(2{\sqrt{\frac{\lambda A_1 \tilde{A}_1}{A^2} }}\right)}{I_{1}\left(2{\sqrt{\frac{\lambda A_1 \tilde{A}_1}{A^2} {}}}\right)}.
\ee
In the next section we will be interested in comparing this result with the strong coupling prediction of super-gravity. For this reason,
we have to expand the above result for large $\lambda$. This can easily be done by recalling that
\be
\frac{I_{k}\left(z\right)}{I_{1}\left(z\right)}=1+O\left(\frac{1}{z}\right).
\ee
Then the correlator in the strong coupling regime becomes
\be
\frac{W_{conn.}}{W_1 W_2}\sim\frac{\lambda}{N^2} \frac{\tilde{A}_1\tilde{A}_2}{A^2}\left[\frac{A_1 A_2}{\tilde{A}_1 \tilde{A}_2} +  2  \left(\sqrt{\frac{A_1 A_2}{\tilde{A}_1 \tilde{A}_2}}\right)^{3}+\cdots\right].
\ee
The first term in the expansion corresponds to the $U(1)$ factor present in $U(N)$ and we shall drop it since it is not generally
considered in the super-gravity analysis. The first non trivial term which can be compared with super-gravity is the second one.


\section{Correlator at strong coupling}

We can also use the AdS/CFT correspondence \cite{Maldacena:1997re} to
compute the correlator of the latitudes at strong coupling, in the
limit where they are well separated compared to their radii, i.e. in
the limit that they migrate to opposite poles of the sphere. In this
limit the correlator is dominated by the exchange of light SUGRA modes
between the two worldsheets describing the Wilson loops at strong
coupling
\cite{Berenstein:1998ij,Semenoff:2004qr,Giombi:2006de,Semenoff:2006am}.

Sometimes, as has been the case for certain chiral primary operators,
two point functions with the Wilson loop can be computed exactly
\cite{Semenoff:2004qr,Semenoff:2006am} in the gauge theory and
succesfully compared at strong coupling to a SUGRA calculation of the
same quantity. Indeed, by taking the ``square-root'' of the
contribution to the correlator of two Wilson loops from a specific
SUGRA mode, the two-point function of the Wilson loop with the
operator dual to that mode is recovered \cite{Berenstein:1998ij}. In
this section we will present a striking agreement between the exchange
of certain such SUGRA modes and the strong-coupling limit of the
QCD$_2$ result (\ref{WWmm}). In order to prove that the QCD$_2$ result
truly captures the correlator at strong coupling, cancellations
between further SUGRA modes will have to be demonstrated. We leave
this to a further publication \cite{toappear}.

\subsection{An intriguing connection}
\label{rpmt}

There appears to be a rather intimate connection between the QCD$_2$ result
presented in section \ref{sec:qcd2} and the two-point functions of
latitude Wilson loops with chiral primary operators built upon the
scalar field $\Phi_3$. In the work \cite{Semenoff:2006am} it was shown
that
\be
\frac{\langle W\, \wt O_J(x) \rangle}{\langle W \rangle}
 = \frac{1}{2 N}\left(\frac{R\sin\theta}{x^2}\right)^J \sqrt{J \lambda} \sin\theta
\frac{I_J\left( \sqrt{\lambda}\, \sin\theta\right)}{I_1\left( \sqrt{\lambda} \sin\theta\right)},
\ee
where $W$ is a latitude Wilson loop at polar angle $\theta$ and
\be\label{phi3}
\wt O_J(x) = \frac{1}{\sqrt{J\lambda}} \Tr\,\left( \Phi_3 + i \Phi_4 \right)^J,
\ee
where $x\gg R\sin\theta$ measures the perpendicular distance between
the operator and the loop. This demonstrates that the matrix model
which yields (\ref{mmVEV}) also captures two-point functions with
those CPO's sharing a minimum amount of SUSY with the latitude Wilson
loop.

Let us look then at the contribution of the $\wt O_J$ to the
correlator of two latitudes, at polar angles $\TO$ and $\TI$, taken
near opposite poles of the sphere to enforce $x\gg R\sin\theta$. Note that
$x = R\cos\TO - R\cos\TI$, we then have
\bsp\label{justOj}
\frac{\langle W_0\,  W_1 \rangle}{\langle W_0 \rangle \langle W_1
  \rangle }\Biggr|_{\wt O_J}
= \frac{\lambda \sin\TO \sin \TI}{4N^2}
\sum_{J=2}^\infty \,J\,
&\left(\frac{\sin\TO \, \sin\TI}{(\cos\TO-\cos\TI)^2}\right)^J\\
&\times \frac{I_J\left( \sqrt{\lambda}\, \sin\theta_0\right)}{I_1\left(
    \sqrt{\lambda} \sin\theta_0\right)}
\frac{I_J\left( \sqrt{\lambda}\, \sin\theta_1\right)}{I_1\left(
    \sqrt{\lambda} \sin\theta_1\right)}.
\end{split}
\ee
This expression is valid strictly at leading order in the large
separation limit. The reason for this is that (\ref{justOj}) ignores
quantum corrections between the propagators joining the operator to
the Wilson loop; this is only valid in the strict large separation
limit as shown in \cite{Pestun:2002mr,Semenoff:2006am}. The expression
(\ref{justOj}) bears a striking resemblance to the QCD$_2$ result
(\ref{WWmm}).  In fact, the only difference lies in the factor in
round parentheses which is risen to the power $J$. However, taking the
large-separation limit of this factor, that difference disappears and
(\ref{justOj}) is {\it exactly} equal to (\ref{WWmm}). Thus the
QCD$_2$ result gives, in the large-separation limit, exactly the
contribution of the exchange of (\ref{phi3}). This agreement is valid
at any value of the coupling, and indeed, in \cite{Semenoff:2006am} it
was shown that at strong coupling the result is recovered from
supergravity.

At leading order in weak coupling, this agreement is puzzling for the
following reason. It is not exactly the operator (\ref{phi3}) which is
present in the latitudes' OPE, since there is no coupling to $\Phi_4$.
Indeed, the calculation of the correlator given in (\ref{OPEA}) shows
that all the operators present in the latitude's OPE (\ref{theop})
participate in the correlator at this order in $\lambda$. It is therefore a
curious coincidence that (\ref{phi3}) produces the same contribution
at weak coupling (i.e. $J=2$) as the true composite operator
(\ref{theop}) present in the actual OPE. Before addressing this issue
further, we present a remarkable strong coupling calculation.

It is interesting to go beyond the strict large-separation limit, and
test the QCD$_2$ result (\ref{WWmm}) to higher orders in the shrinking
radii of the two latitudes. It turns out that at strong coupling, the
associated SUGRA calculation giving this information is tractable. In
keeping with the intriguing connection between the contribution of
(\ref{phi3}) to the correlator and the QCD$_2$ result, we begin by
computing the exchange of the SUGRA modes dual to (\ref{phi3}) in an
expansion about small latitude radii $\TO$ and $\TI$ (where the polar
angle of the latitude at the south pole is given by $\pi-\TI$).

The supergravity modes dual to (\ref{phi3}) are fluctuations of
the RR 5-form as well as the spacetime metric. They are by now very
well known, and details can be found in
\cite{Berenstein:1998ij}\cite{Lee:1998bxa}\cite{Kim:1985ez}\cite{Semenoff:2004qr}\cite{Giombi:2006de}.
The fluctuations of the metric are
\bea\label{fluct} \d g_{\m \n} &=&
\left[-\frac{6\,J}{5}\,g_{\m \n} + \frac{4}{J+1} \, D_{(\m}
  D_{\n)} \right] \,s^J(x)\,Y^J(\Omega),\cr \d g_{\a\b } &=& 2\,J\,g_{\a\b
} \,s^J(x)\,Y_J(\Omega),
\eea
where $\m,\n$ are $AdS_5$ and $\a,\b$ are $S^5$ indices. The symbol
$x$ indicates coordinates on $AdS_5$ and $\Omega$ coordinates on the
$S^5$. The bulk-to-bulk scalar propagator for the field $s^J(x)$
is\footnote{See
  \cite{Berenstein:1998ij}\cite{Lee:1998bxa}\cite{Kim:1985ez}\cite{Semenoff:2004qr}\cite{Giombi:2006de}
  for the definitions of $\a_0$ and $B_J$.}
\be
P(x,\bar x) = \frac{\alpha_0}{B_J} \, W^J \, {_2}F_1 (\,J, \,J - 3/2, \,2J - 3;
\,-4W)
\ee
where in an $AdS_5$ given by $ds^2 = (dx_0^2+dx_i^2)/x_0^2$,
$W=x_0\bar x_0/((x_0-\bar x_0)^2+(x_i-\bar x_i)^2)$. The full details
of the calculation will be presented in \cite{toappear}, however it is
essentially that found in \cite{Semenoff:2006am}. There, the strict
large-separation limit was employed by setting the hypergeometric
function to 1. Here we keep higher terms in the expansion. The results are as follows
\bsp
&J = 2~: \qquad  \frac{\langle W(x)\, W(\bar x) \rangle  }{ \langle W(x)\rangle\langle
  W(\bar x)\rangle}= \frac{\lambda}{8 N^2} \Biggl[
\frac{\TO^3 \,\TI^3}{2^2} + \frac{\TO^3\,\TI^7 + \TO^7\,\TI^3}{5\cdot 3\cdot 2^6}
+ \frac{\TO^5\,\TI^5}{2^6} + \frac{\TO^3\,\TI^9 + \TO^9\,\TI^3}{7\cdot
  3^3\cdot 2^6}\\
&\qquad\qquad\qquad\qquad\qquad\qquad\qquad\qquad
+ \frac{\TO^5\,\TI^7 + \TO^7\,\TI^5}{3\cdot 2^7}
+  \frac{\TO^6\,\TI^6}{5^2\cdot 3} -
\frac{\TO^7\,\TI^6+\TO^6\,\TI^7}{5\cdot 3 \cdot 2^3} + {\cal
  O}(\theta^{14})\Biggr],\\
&J = 3~: \qquad  \frac{\langle W(x)\, W(\bar x) \rangle  }{ \langle W(x)\rangle\langle
  W(\bar x)\rangle}= \frac{\lambda}{32 N^2} \Biggl[
\frac{3\,\TO^4 \,\TI^4}{8} + \frac{\TO^4\,\TI^6 + \TO^6\,\TI^4}{2^5}
 + \frac{3\,(\TO^4\,\TI^8 + \TO^8\,\TI^4)}{5\cdot 2^7}\\
&\quad
+ \frac{5\,\TO^6\,\TI^6}{3\cdot 2^6} + \frac{3^3\,\TO^7\,\TI^7}{7^2\cdot 5^2}
+ \frac{(\TO^6\,\TI^8 + \TO^8\,\TI^6)}{5\cdot 2^5}
+ \frac{23\,(\TO^4\,\TI^{10} + \TO^{10}\,\TI^4)}{7\cdot 5\cdot 3^3\cdot 2^7}
- \frac{3^2\,(\TO^7\,\TI^8 + \TO^8\,\TI^7)}{7\cdot 5\cdot 2^5}
+ {\cal O}(\theta^{16})\Biggr],\\
&J = 4~: \qquad  \frac{\langle W(x)\, W(\bar x) \rangle  }{ \langle W(x)\rangle\langle
  W(\bar x)\rangle}= \frac{\lambda}{256 N^2} \Biggl[
\TO^5 \,\TI^5 + \frac{\TO^5\,\TI^7 + \TO^7\,\TI^5}{3\cdot2}
 + \frac{\TO^5\,\TI^9 + \TO^9\,\TI^5}{3^2\cdot 2^2}
+ \frac{13\,\TO^7\,\TI^7}{3^2\cdot 2^4}\\
&\qquad\qquad\qquad\qquad\qquad\qquad\qquad\qquad
+ {\cal O}(\theta^{16})\Biggr].
\end{split}
\ee
\ni The QCD$_2$ result (\ref{WWmm}) in the large $\lambda$ limit is
\be
\frac{\langle W(x)\, W(\bar x) \rangle  }{ \langle W(x)\rangle\langle
  W(\bar x)\rangle} \Biggr|_{QCD_2}=\frac{\lambda \sin\TO \, \sin \TI}{4N^2} \sum_{J=1}^\infty
J\,\tan^J\frac{\TO}{2} \, \tan^J \frac{\TI}{2} .
\ee
Ignoring $J=1$, we may expand in $\theta$ order-by-order in $J$:
\bsp\label{qcd2}
&J = 2~: \qquad  \frac{\langle W(x)\, W(\bar x) \rangle  }{ \langle W(x)\rangle\langle
  W(\bar x)\rangle}\Biggr|_{QCD_2} = \frac{\lambda}{8 N^2} \Biggl[
\frac{\TO^3 \,\TI^3}{2^2} + \frac{\TO^3\,\TI^7 + \TO^7\,\TI^3}{5\cdot 3\cdot 2^6}
+ \frac{\TO^3\,\TI^9 + \TO^9\,\TI^3}{7\cdot
  3^3\cdot 2^6}+ {\cal O}(\theta^{14})\Biggr],\\
&J = 3~: \qquad  \frac{\langle W(x)\, W(\bar x) \rangle  }{ \langle W(x)\rangle\langle
  W(\bar x)\rangle}\Biggr|_{QCD_2}= \frac{\lambda}{32 N^2} \Biggl[
\frac{3\,\TO^4 \,\TI^4}{8} + \frac{\TO^4\,\TI^6 + \TO^6\,\TI^4}{2^5}
 + \frac{3\,(\TO^4\,\TI^8 + \TO^8\,\TI^4)}{5\cdot 2^7}
\\
&\qquad\qquad\qquad\qquad\qquad\qquad + \frac{\TO^6\,\TI^6}{3\cdot 2^7}
+ \frac{(\TO^6\,\TI^8 + \TO^8\,\TI^6)}{5\cdot 2^{9}}
+ \frac{23\,(\TO^4\,\TI^{10} + \TO^{10}\,\TI^4)}{7\cdot 5\cdot 3^3\cdot 2^7}
+ {\cal O}(\theta^{16})\Biggr],\\
&J = 4~: \qquad  \frac{\langle W(x)\, W(\bar x) \rangle  }{ \langle W(x)\rangle\langle
  W(\bar x)\rangle}\Biggr|_{QCD_2}= \frac{\lambda}{256 N^2} \Biggl[
\TO^5 \,\TI^5 + \frac{\TO^5\,\TI^7 + \TO^7\,\TI^5}{3\cdot2}
 + \frac{\TO^5\,\TI^9 + \TO^9\,\TI^5}{3^2\cdot 2^2}
+ \frac{\TO^7\,\TI^7}{3^2\cdot 2^2}\\
&\qquad\qquad\qquad\qquad\qquad\qquad\qquad\qquad\qquad\qquad
+ {\cal O}(\theta^{16})\Biggr].
\end{split}
\ee
There is a remarkable matching of highly non-trivial terms between
these two calculations! The difference between the two calculations
sets-in quite late
\bsp
&\left( \text{SUGRA} - \text{QCD}_2\right)_{J=2} = \frac{\lambda}{8 N^2} \Biggl[ \frac{\TO^5\,\TI^5}{2^6}
+ \frac{\TO^5\,\TI^7 + \TO^7\,\TI^5}{3\cdot 2^7}
+  \frac{\TO^6\,\TI^6}{5^2\cdot 3} -
\frac{\TO^7\,\TI^6+\TO^6\,\TI^7}{5\cdot 3 \cdot 2^3} + {\cal
  O}(\theta^{14})\Biggr],\\
&\left( \text{SUGRA} - \text{QCD}_2\right)_{J=3}\!=\! \frac{\lambda}{32 N^2}\!
\Biggl[ \frac{3\,\TO^6\,\TI^6}{2^7}
+ \frac{3\,(\TO^6\,\TI^8 + \TO^8\,\TI^6)}{2^{9}}
 + \frac{3^3\,\TO^7\,\TI^7}{7^2\cdot 5^2}
- \frac{3^2\,(\TO^7\,\TI^8 + \TO^8\,\TI^7)}{7\cdot 5\cdot 2^5}\!
+ \!{\cal O}(\theta^{16})\!\Biggr]\!,\\
&\left( \text{SUGRA} - \text{QCD}_2\right)_{J=4} = \frac{\lambda}{256 N^2}
\Biggl[ \frac{\TO^7\,\TI^7}{2^4}
+ {\cal O}(\theta^{16})\Biggr].\\
\end{split}
\ee
Although we have considered values of $J$ up to $J=4$, we expect a
similar pattern for arbitrary $J$.

\subsection{Other modes}
\label{sec:scd}

The remarkable agreement displayed in the previous section does {\it
  not} prove that the QCD$_2$ result captures the correlator of the
latitudes at strong coupling. Beyond the issue of the discrepancy at
order $\TO^5\TI^5$, the catch is that the SUGRA spectrum contains two
other modes which couple to the string worldsheets and also produce
$\TO^3\TI^3$ terms, thereby potentially spoiling the agreement with
the QCD$_2$ result. These are the Kaluza-Klein modes of the NS-NS
B-field of type-IIB supergravity, and have been described in
\cite{Kim:1985ez}, c.f. their equation (2.48) and what follows it.
There is a fluctuation of the B-field with both legs in the $S^5$
which is described by a scalar of mass-squared $-3$ (corresponding to
a gauge theory operator of protected dimension 3) given by
\be\label{NSNSmodes}
\d B_{\a\b} = a^{k}_-(x)\, Y_{[\a\b]}^{k,-}(\Omega),\qquad m_{a^k_-}^2 = k^2-4,
\ee
with $k=1$.
%
%
There is also the fluctuation of the B-field with both legs in the
$AdS_5$ portion of the geometry $\d B_{\m\n}$, which has been
discussed in \cite{Arutyunov:1998xt}. It has the Kaluza-Klein
expansion
\be
\d B_{\m\n} = a^k_{\m\n}(x) \, Y^k(\Omega),\qquad m_{a^k_{\m\n}}^2 = k^2,
\ee
and the leading $k=1$ harmonic corresponds to the following protected
dimension 3 operator (where $A,B$ are SU(4) indices)
\be\label{obmn}
2 i\, \Phi^{AB} F^+_{\m\n}+
\bar \psi^A  \s_{\m\n} \bar \psi^B.
\ee
These contributions must cancel out if the QCD$_2$ result is to hold.
Beyond these modes, there are also fluctuations of the dilaton,
massless vector, and massless tensor which provide contributions which
lead as $\TO^4 \TI^4$ and must therefore also find a way to cancel
each other, should the QCD$_2$ result truly describe the correlator at
strong coupling. Indeed this is the reflection at strong coupling of
the curiosity of the fact that the operators of classical dimension 3
and 4 contributing to the correlator at weak coupling seem to have the
same effect as replacement by (\ref{phi3}) (with $J=2$). The full
calculation of these SUGRA modes, and the question of whether or not
they cancel, will be explored in a companion publication
\cite{toappear}.

The matrix model result (\ref{mmVEV}) contains a rescaled coupling
constant $\lambda'=\lambda \sin^2\theta$. The two point function of the latitude
with the CPO (\ref{phi3}) leads as $\lambda'$ but ends-up as $\sqrt{\lambda'}$
at strong coupling. This explains why in the OPE the operator
$\Tr\,\Phi_3^2$ is weighted by $\theta^4$ but ends-up contributing as
$\theta^3$ at strong coupling. The first descendent of this operator
 appearing in the OPE of the latitude is
$\Tr\,\Phi_3
\partial_3 \Phi_3$ and comes with weight $\theta^6$, thus one would
expect its contribution at strong coupling to be $\theta^5$,
potentially explaining why the discrepancy between the QCD$_2$ result
and the contribution from CPO's built on $\Phi_3$ sets-in at order
$\TO^5 \TI^5$.

\section*{Acknowledgments}

L.G and D.S. thanks Giulio Bonelli and Alessandro Tanzini for discussions. L.G.
and D.S. thanks the Galileo Galilei Institute for hospitality and support.
D.Y. thanks Nadav Drukker, Jan Plefka, Johannes Henn, Harald Dorn, and
George Jorjadze for discussions. D.Y. acknowledges the support of the
Natural Sciences and Engineering Research Council of Canada (NSERC) in
the form of a Postdoctoral Fellowship, and also support from the
Volkswagen Foundation.

\appendix
\setcounter{equation}{0}
\renewcommand{\theequation}{\thesection.\arabic{equation}}
\section{Appendix}
\label{infrabehav}
\noindent
The  integral $\mathcal{I}_1$ defined in \eqref{II1}, for example, was computed in \cite{Bassetto:2008yf}
and a useful representation for the final result is
\be
\mathcal{I}_1(x_1,x_2,x_3)=
\frac{ 1 }{64\pi^{4}}
\!\!\!\int_0^1\!\!\!\!
 d\alpha \frac{1}{
 ({y}-\alpha{x})^2}\log\left[\frac{\alpha[({x}-{y})^2-{y}^2]+
 {y}^2}{\alpha(1-\alpha){x}^2}\right],
\ee
where $x=x_1-x_2$ and $y=x_3-x_2$.
The only logarithmic behavior in this integral arises when $x_1$ and $x_2$ approach the same point $x_0$ (namely $|x|\to 0$), and is given by
\be
\label{I1}
\mathcal{I}_1^{sing.}=-\frac{ 1 }{64\pi^{4}}
\!\!\!\int_0^1\!\!\!\!
 d\alpha \frac{1}{({y}-\alpha{x})^2}\log{x}^2=-\frac{ 1 }{64\pi^{4}}
\!\!\!\int_0^1\!\!\!\!
 d\alpha \frac{\log (x_1-x_2)^2}{({(x_3-x_2)}-\alpha{(x_1-x_2)})^2}.
\ee
Next we consider the integral
\be
\mathcal{I}^{(4)}(x_1,x_2,x_3,x_4)=\frac{1}{(4\pi^2)^4}
\int \frac{d^{4} z}{(x_1-z)^2(x_2-z)^2 (y_1-z)^2 (y_2-z)^2}.
\ee
It is well-known that this integral can be computed in terms of $\mathcal{I}_{1}$ \cite{Drummond:2006rz}. In fact if
we define
\be
\bar{x}_1^\mu=\frac{(x_1-y_2)^\mu}{(x_1-y_2)^2},\ \ \ \bar{x}_2^\mu=\frac{(x_2-y_2)^\mu}{(x_2-y_2)^2},\ \ \ \
\bar{x}_3^\mu=\frac{(y_1-y_2)^\mu}{(y_1-y_2)^2},
\ee
we find
\be
\mathcal{I}^{(4)}(x_1,x_2,x_3,x_4)=\frac{\bar{x}_1^2\bar{x}_2^2\bar{x}_3^2}{(4\pi^2)^4}
\int \frac{d^{4} z}{(\bar{x}_1-z)^2(\bar{x}_2-z)^2 (\bar{x}_3-z)^2}=
\frac{\bar{x}_1^2\bar{x}_2^2\bar{x}_3^2}{4\pi^2} \mathcal{I}_1(\bar{x}_1-\bar{x}_2, {\bar{x}_3}-\bar{x}_2).
\ee
Then
\begin{align}
&\mathcal{I}^{(4) sing.}(x_1,x_2,y_1,y_2)=-\frac{{ \log(x_1-x_2)^2 }}{256\pi^6}\times\nonumber\\
&\times
\!\int_0^1\!\!
  \frac{d\alpha}{\mbox{\small $(1-\alpha){( y_1- x_2)^2(y_2 -x_1)^2}-\alpha(1-\alpha) {( x_1- x_2)^2(y_1-y_2)^2}+
{\alpha( y_1- x_1)^2(y_2-x_2)^2}$}}.
\end{align}
\noindent
For our goals, the most convenient way  to compute the integral $\mathcal{S}^\mu$ defined in \eqref{AA} is to use
the technique of \cite{Davydychev:1991va}, which allows us to reduce  the tensor integrals to scalar integrals in higher
space-time dimensions.  We shall perform this reduction in  $2\omega$ dimensions and for arbitrary powers of
the denominators.  The final result is very nice and compact
\be
\prod_{i=1}^4\frac{\Gamma(a_i)}{4\pi^{a_i+1}}\int  \frac{w^\mu d^{2\omega} w}{((x_1-w)^2)^{a_1} ((x_2-w)^2)^{a_2}((x_3-w)^2)^{a_3}((x_4-w)^2)^{a_4}}=\sum_{j=1}^4  x_j^\mu \mathfrak{S}(\omega+1;a_i+\delta_{ij})
\ee
where
\be
\mathfrak{S}^{(2\omega)}(\omega;a_i)=\prod_{i=1}^4\frac{\Gamma(a_i)}{4\pi^{a_i+1}}\int  \frac{d^{2\omega} w}{((x_1-w)^2)^{a_1} ((x_2-w)^2)^{a_2}((x_3-w)^2)^{a_3}((x_4-w)^2)^{a_4}}.
\ee
In computing $\mathbf{A_1}$  we  also need the derivative with respect to $x_2^\nu$ of the above expression. After some manipulation
this derivative can be arranged as follows
\be
\label{B18}
\begin{split}
&\prod_{i=1}^4\frac{\Gamma(a_i)}{4\pi^{a_i+1}}\frac{\partial}{\partial x_2^\nu}\int  \frac{w^\mu d^{2\omega} w}{((x_1-w)^2)^{a_1} ((x_2-w)^2)^{a_2}((x_3-w)^2)^{a_3}((x_4-w)^2)^{a_4}}=\\
&=\delta^{\mu\nu} \mathfrak{S}(\omega+1;a_i+\delta_{i2})+2\pi
\sum_{k=1}^4\sum_{j=1}^4  x_j^\mu (x_k-x_2)^\nu \mathfrak{S}(\omega+2;a_i+\delta_{ij}+\delta_{i2}+\delta_{ki}).
\end{split}
\ee
Finally, the only other ingredient necessary for our calculation is the behavior of the integral $\mathcal{S}(2\omega; a_i)$ when
$x_1$ and $x_2$ approach the same point $x_0$.
\be
\begin{split}
\mathcal{S}(2\omega; a_i)
=&\frac{}{}
\frac{ \Gamma \left(\omega -a_1\right) \Gamma \left(\omega -a_2\right) \Gamma(a_3)\Gamma(a_4)\Gamma \left(
   a_1+a_2-\omega\right)}{256\pi^{\mbox{\tiny $\displaystyle{\sum_{i=1}^4}$} a_i+4-\omega}\Gamma \left(2 \omega -a_1-a_2\right)}
\frac{(({{x}_1}-{{x}_2})^2)^{\left(\omega -a_1-a_2\right)}}{(({x}_3)^2)^{a_3}(({x}_4)^2)^{a_4}}
\times\\
&\times\left[1+ 2  \left(a_3\frac{{x}_3}{{x}_3^2}+a_4\frac{{x}_4}{{x}_4^2}\right)\cdot\left(({x}_2-x_0)+\frac{\omega-a_2}{2\omega-a_1-a_2} ({x}_1-{x}_2)\right)+O(({x}_1-{x}_2)^2)\right].
\end{split}
\ee

\end{document}